\documentclass[10pt,journal,compsoc]{IEEEtran}
\newif\ifpeerreview

\peerreviewfalse

\usepackage{graphicx}
\usepackage{amsmath}

\usepackage{amssymb}
\usepackage{subcaption}
\usepackage{enumitem}
\usepackage{animate}
\usepackage[numbers, sort&compress]{natbib}
\usepackage{url}
\usepackage{xcolor}
\usepackage{capt-of}
\usepackage{etoolbox}

\setlist{leftmargin=*}

\usepackage[ruled]{algorithm2e} 

\SetAlFnt{\small}
\SetAlCapFnt{\small}
\SetAlCapNameFnt{\small}
\SetAlCapHSkip{0pt}

\newcommand{\bpara}[1]{{\flushleft \textbf{#1}}}

\usepackage[switch]{lineno}

\newcommand{\paperID}{0005}

\title{SASSI --- Super-Pixelated Adaptive Spatio-Spectral Imaging}

\author{Vishwanath~Saragadam$^\dagger$, Michael~DeZeeuw$^\ddag$, Richard~Baraniuk$^\dagger$,\\ Ashok~Veeraraghavan$^\dagger$, and Aswin~Sankaranarayanan$^\ddag$\\
$^\dagger$ECE Department, Rice University, Houston TX\\
$^\ddag$ECE Department, Carnegie Mellon University, Pittsburgh PA}

\begin{document}

\IEEEtitleabstractindextext{%
\begin{abstract}
We introduce a novel video-rate  hyperspectral imager with high spatial, and temporal resolutions.
Our key hypothesis is that spectral profiles of pixels in a super-pixel of an oversegmented image tend to be very similar.
Hence, a scene-adaptive spatial sampling of an hyperspectral scene, guided by its super-pixel segmented image, is capable of obtaining high-quality reconstructions.
To achieve this, we acquire an RGB image of the scene, compute its super-pixels, from which we  generate a spatial mask of locations where we measure high-resolution spectrum.
The  hyperspectral image is subsequently estimated by fusing the RGB image and the spectral measurements  using a  learnable guided filtering approach.
%
%
Due to low computational complexity of the superpixel estimation step, our setup can capture hyperspectral images of the scenes with little overhead over traditional snapshot hyperspectral cameras, but with significantly higher spatial and spectral resolutions.
%
%
We validate the proposed technique with extensive simulations as well as a  lab prototype that measures hyperspectral video at a spatial resolution of $600 \times 900$  pixels,  at a spectral resolution of 10 nm over visible wavebands, and achieving a frame rate at $18$fps.
\end{abstract}

}

\twocolumn[{%
	\renewcommand\twocolumn[1][]{#1}%
	\maketitle
	\vspace{-4em}
	\begin{center}
		\centering
		\begin{tabular}{c}
			\animategraphics[width=\textwidth,loop]{5}{figures/matroshka_video/stack/im}{1}{50}
		\end{tabular}
		\includegraphics[width=0.9\textwidth]{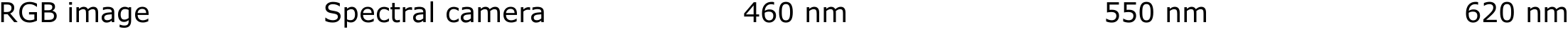}
		\captionof{figure}{\textbf{High resolution video rate hyperspectral imaging.} We propose a novel hyperspectral camera that is capable of capturing high spatial and spectral resolution images at video rate. We achieve this by a sparse, scene-adaptive spectral sampling, and then fusing it with an auxiliary RGB image. Please use Adobe Reader to view the animation.}
		\label{fig:teaser}
			\vspace{2em}
	\end{center}
}]


\IEEEraisesectionheading{
  \section{Introduction}\label{section:introduction}
}
\IEEEPARstart{O}{ne} of the classic approaches for plenoptic imaging is to sacrifice  spatial resolution of a sensor to obtain resolution in other dimensions of light.
This principle has been used for sensing color with filter arrays \cite{bayer1976color},  polarization state with per-pixel polarizers  \cite{gruev2010ccd}, light fields with angle sensitive pixels \cite{hirsch2014switchable} and integral imagers \cite{ng2005light}, high-speed imaging with staggered pixel exposures \cite{bub2010temporal}, and high dynamic range with per-pixel neutral density filters \cite{nayar2000high}.
This trade-off is succinctly encapsulated in the concept of assorted pixels \cite{narasimhan2005enhancing} where the pixels in the sensor  also sample along non-spatial dimensions of the incident light --- be it time, spectrum,  angle,  dynamic range or combinations thereof.
The main challenge in sensing with assorted pixels has been the steep loss in spatial resolution, especially when we need to allocate a large portion of samples to other dimensions.

This paper provides a design template for novel hyperspectral cameras that work in the spirit of assorted pixels, i.e., trading  spatial resolution of the sensor to obtain high spectral resolutions.
The key differentiating aspect of our design is that it performs a \textit{scene-adaptive} tiling of the spatio-spectral voxels onto the image sensor, which is in sharp contrast to prior work where the measurements are  non-adaptive.
Our approach relies on an assumption that, for most scenes, the spectrum of pixels is extremely similar over small spatial neighborhoods that do not cross texture and material boundaries, such as superpixels.
Hence, if we knew the regions with homogeneous spectrum a priori, then we could sample one or more spectral profile for each region and propagate it to the remaining pixels, which would avoid blurring of the measurements as well as loss of spatial resolution.
We refer to our approach as  Super-pixelated  Adaptive Spatio-Spectral Imager (SASSI).
%

%
SASSI  relies on two components: a \textit{guide RGB camera}, that is used to determine the  spatial locations where we sample spectrum; and,  a \textit{spatio-spectral sampler}, that is capable of measuring the high-resolution spectrum at  chosen spatial locations by smearing their spectrum in space.
%
%
The spatio-spectral sampler is  identical to that of the single disperser coded aperture snapshot spectral imager (CASSI) architecture \cite{wagadarikar2008single}, except that we have a  programmable mask in the form of the SLM.
Both components share the same view point and are time synchronized.

To acquire the hyperspectral image (HSI) at a time instant $t$, we first 
perform super-pixel segmentation on the guide image acquired at the previous time instant, say $t-1$, to obtain a clustering of compact regions that have similar RGB intensities; we use this as a proxy for the material map of the scene.
We subsequently  select pixels such that they satisfy two key criteria: first, pixels are chosen sufficiently far apart from each other so that their spectral spreads in the spatio-spectral sampler do not overlap;  and second, the number of  super-pixels selected is maximized along with the total sensor pixel count in the spatio-spectral sampler.
We now acquire measurements at time $t$, with the adaptively designed mask displayed on the SLM.
%
Finally, we can estimate the HSI by fusing the RGB image and the spatio-spectral samples acquired at time $t$.
Since all measurements being fused are captured simultaneously, in a time-synchronized way, we avoid the need for registration of images across time instants.
%
%
%


{{\flushleft \textbf{Contributions.}}}
We propose SASSI, an adaptive HSI sensing approach, and make the following contributions.
\begin{itemize}[leftmargin=*]
\item \textit{Optical design.} We provide a compact, light efficient design for the SASSI imager.
%
\item \textit{Reconstruction of the HSI.} We propose two recovery approaches applicable to a variety of scenarios. The first one is a rank-1 reconstruction where each pixel within the superpixel is assigned a scaled version of the measured spectrum at the centroid. The second one is a per-wavelength neural network based reconstruction that relies on learnable guided filtering. 
%
%
\item \textit{Lab prototype.} We build a prototype of the SASSI system that is capable of acquiring HSIs with a spatial resolution of $698 \times 931$ over a visible waveband $400 - 700$ nm at a spectral resolution of 10 nm, operating at 18 fps.
\end{itemize}

We  provide extensive evaluation  using simulations and real data captured using the  prototype.
Our results come with a new HSI dataset with $50$ scenes, including some microscopic samples, which we will release publicly.
%
%

{{\flushleft \textbf{Limitations.}}}
The limitations of the SASSI architecture are three fold.
\begin{itemize}[leftmargin=*]
\item \textit{Thin spatial structures and clutter.} Our setup requires that sampling locations not be close to each other, which precludes measurement of highly complex spatial structures.
%
%
\item \textit{Scenes with rapid motion.}  
Our approach fuses RGB and spectral measurements from the same time instance; however, rapid motion between frames may create sampling locations that are off from the desired pattern.
%
\item \textit{Metamerism.}  
For scenes with metamerism, we run the risk of violating the assumption that the spectrum remains largely the same within each super-pixel.
\end{itemize}

\section{Prior Work} \label{section:prior_work}
We review some of the key prior art in hyperspectral imaging and adaptive sensing.

\subsection{Hyperspectral camera designs}
The SASSI architecture is closely related to many existing  designs for hyperspectral imaging; we discuss them next.

\begin{figure}[!tt]
	\centering
	\includegraphics[width=\columnwidth]{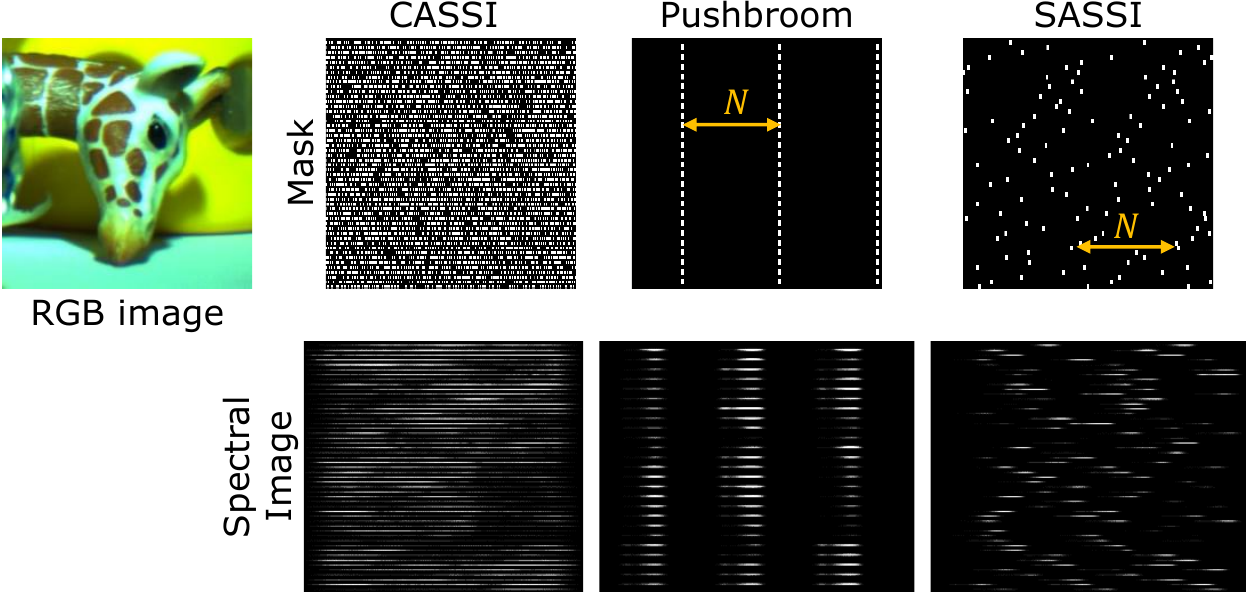}
	\caption{\textbf{Various sampling schemes.} CASSI-type cameras sense with a dense spatial pattern, requiring spectral demultiplexing. On the other hand, pushbroom cameras do not require demultiplexing but require long capture durations. SASSI provides a unique tradeoff. The sampling pattern is sparse enough to avoid any spectral multiplexing, while requiring only a single, scene-adaptive spectral measurement.}
	\label{fig:sampling_compare}
\end{figure}

{{\flushleft \textbf{Classical hyperspectral cameras.}}}
Hyperspectral images capture information as a function of both space and wavelength, and can be represented as a 3D volume, $H(x, y, \lambda)$.
Spectral profiles are often unique identifiers of materials and illuminants and have been used in several scientific \cite{cloutis1996review,lichtman2005fluorescence,zhi2019multispectral} and computer vision \cite{goel2015hypercam,saragadam2019programmable,hui2018illuminant} applications.
%
Traditional cameras either one wavelength at a time, or all wavelengths of one row of the image at a time, both of which require long exposure times.
%
%
%
The classical pushbroom camera samples a single or multiple spatial columns of the HSI by smearing their spectral profiles onto the spatial sensor; this can be seen in  Figure \ref{fig:sampling_compare}.
Both pushbroom and the proposed SASSI avoid the spectral streaks from overlapping with each other, and hence, they both provide a sub-sampled Nyquist sampling of the HSI. The key difference lies in the uniform non-adaptive sampling in pushbroom versus the non-uniform adaptive sampling in SASSI.

%


{{\flushleft \textbf{Compressive sensing}}} refers to a class of techniques that senses a signal from a set of random non-adaptive linear measurements, whose cardinality is smaller than the signal dimension.
%
%
%
The CASSI camera and its variants \cite{wagadarikar2008single,kittle2010multiframe,lin2014spatial,arce2014compressive,arguello2013rank,rueda2016compressive,rueda2017high,deepcassi2017} are examples of compressive imagers where a single spatio-spectrally multiplexed image is demultiplexed into a full-resolution HSI using various signal priors
%
Such cameras lead to severe loss of spatial resolution, and  require solving complex optimization problems.
%
%
SASSI can be interpreted as an extension of the CASSI architecture.
In addition to being  adaptive, our method avoids multiplexing of spectrum across pixels. This enables reconstructions that are both accurate and faster.
Figure \ref{fig:sampling_compare} shows the sampling masks for the CASSI system; the higher density of openings in CASSI leads to multiplexing of spectral measurements from different spatial locations, which is a key difference to the proposed technique.

{{\flushleft \textbf{Hybrid cameras.}}}
Hybrid cameras  fuse a low-resolution hyperspectral imager and a high-resolution RGB camera to obtain a high spatial and spectral camera; this process is also referred to panchromatic sharpening \cite{loncan2015hyperspectral,kawakami2011high}.
Cao et al.\ \cite{cao2011high} use a CASSI system in place of a low-resolution hyperspectral camera.
Instead of obtaining a random and dense spatio-spectral multiplexing, they rely on a uniform  sparse mask  with a transmission efficiency  $<0.03$\% and then propagate the sparse set of spectral to all other locations by fusing it with an RGB image.
%
%
However, a uniform sampling strategy is not suitable  for  scenes that have small, irregular objects.
%

%
%

\subsection{Adaptive sampling strategies}
Instead of fixing the sampling scheme across all scenes, it is intuitive that an adaptive  sampling tuned to the specifics of an individual scene can provide higher accuracy \cite{kauvar2015adaptive,o2010optical,saragadam2018krism}.
In the context of snapshot hyperspectral imaging, Feng et al.\ \cite{fang2014texture} and Ma et al.\ \cite{ma2014content} showed that an RGB image can be used to tailor the mask for CASSI, thereby creating a content-adaptive hyperspectral sampling strategy.
This is a promising approach to snapshot hyperspectral imaging, and in many ways, is a key motivation for our own work.

There are some important differences  between SASSI and those mentioned above.
%
%
%
Ma et al.\ \cite{ma2014content}  rely on temporal correction of errors, as well as a simple modification of bilateral filtering in the reconstruction procedure, which  requires a very accurate estimation of correspondences across time-frames --- a task that is often difficult in the absence of texture in the scene.
%
%
SASSI instead relies on RGB image and spectral information from the \textit{same} frame, thereby not requiring any  image alignment.
%
%
Further, we rely on a learned guided filtering based reconstruction technique that can recover HSIs with high accuracy.
%
%
%

\subsection{Super-pixelation of images}
%
Super pixels provide an over-segmentation of an image into  homogeneous regions \cite{achanta2012slic,felzenszwalb2004efficient,levinshtein2009turbopixels}, and have been used extensively in vision problems.
%
%
Super pixellation is computationally light-weight as compared to  approaches such as semantic segmentation since it primarily relies on spatial proximity and RGB intensity to estimate segment boundaries. 
%
%
We use  the Simple Linear Iterative Clustering (SLIC) algorithm \cite{achanta2012slic} --- a fast light-weight technique that is especially suitable for real-time applications.
%
%
%

\begin{figure}[!tt]
	\centering
	\includegraphics[width=\columnwidth]{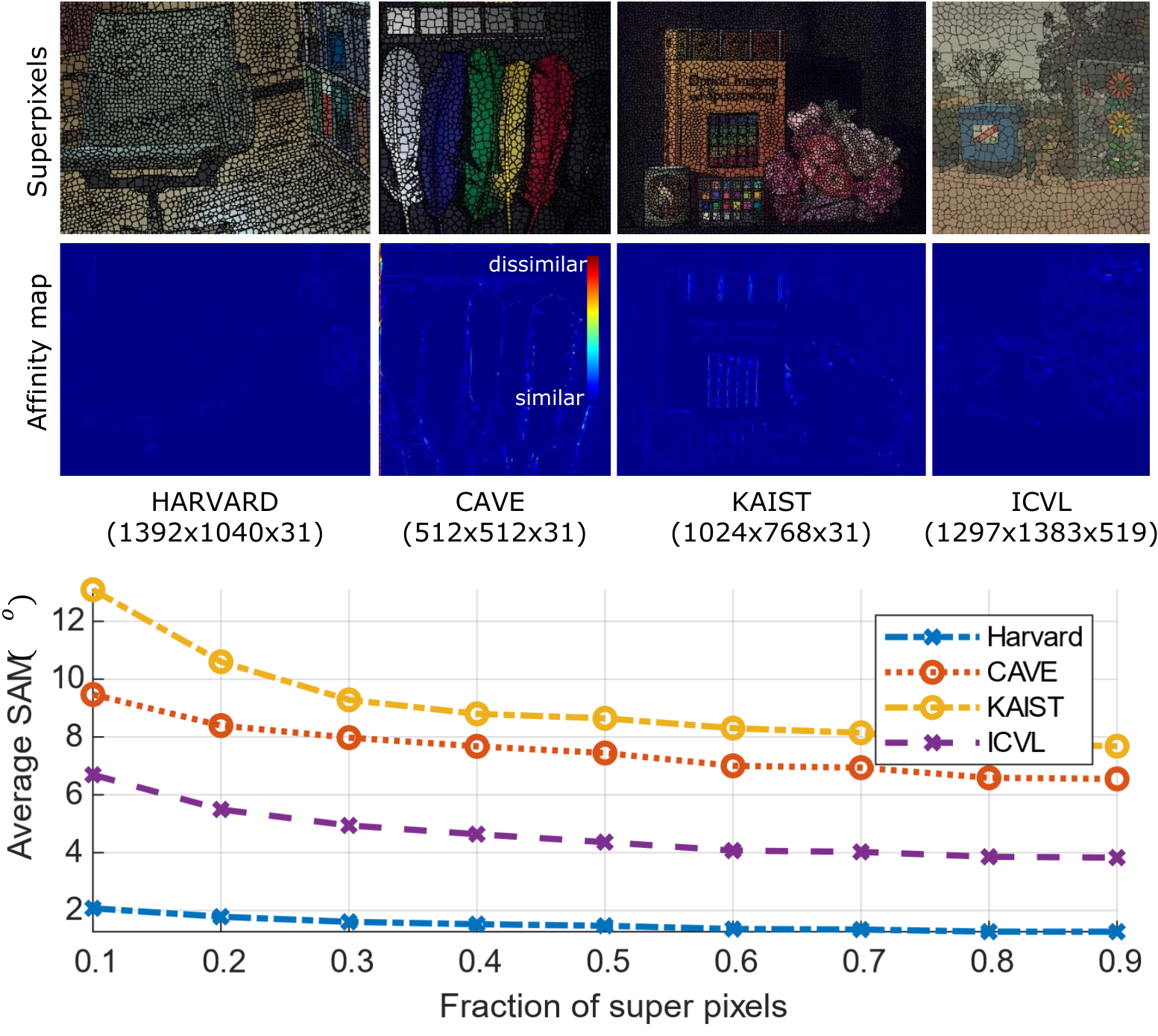}
	\caption{ \textbf{Homogeneity of super-pixels.} We hypothesize that super-pixels represent homogeneous regions of spectra. To verify this, we estimated similarity between spectral profile at one location within a super-pixel and all its other members for commonly available datasets. We observe that spectral profiles inside a super pixel are highly correlated, evident from the the very small Spectral Angular Mapping (SAM) value.}
	\label{fig:affinity}
\end{figure}

The central hypothesis of this paper is that spectral profiles  within a super pixel tend to be similar.
We verify this on  HSIs from several  datasets \cite{yasuma2010generalized,deepcassi2017,chakrabarti2011statistics,arad_and_ben_shahar_2016_ECCV}
by rendering RGB images, and estimating superpixels.
Then we estimate the similarity between spectrum at one location and all its neighbors within its superpixel in terms of spectral angular similarity (SAM) (see Fig. \ref{fig:affinity}).
%
%
The average SAM value for any given HSI was less than $10^\circ$ --- corresponding approximately to 30 dB --- which states that super pixels are a good way of separating images into homogeneous regions.
Intuitively, it should then suffice to sample at one location within each super pixel
%
%
Such a strategy reduces complexity of sampling by providing a single shot hyperspectral imager with extremely fast reconstruction.
%

\section{The SASSI Camera} \label{section:proposed}

\begin{figure}[!tt]
	\centering
	\includegraphics[width=\columnwidth]{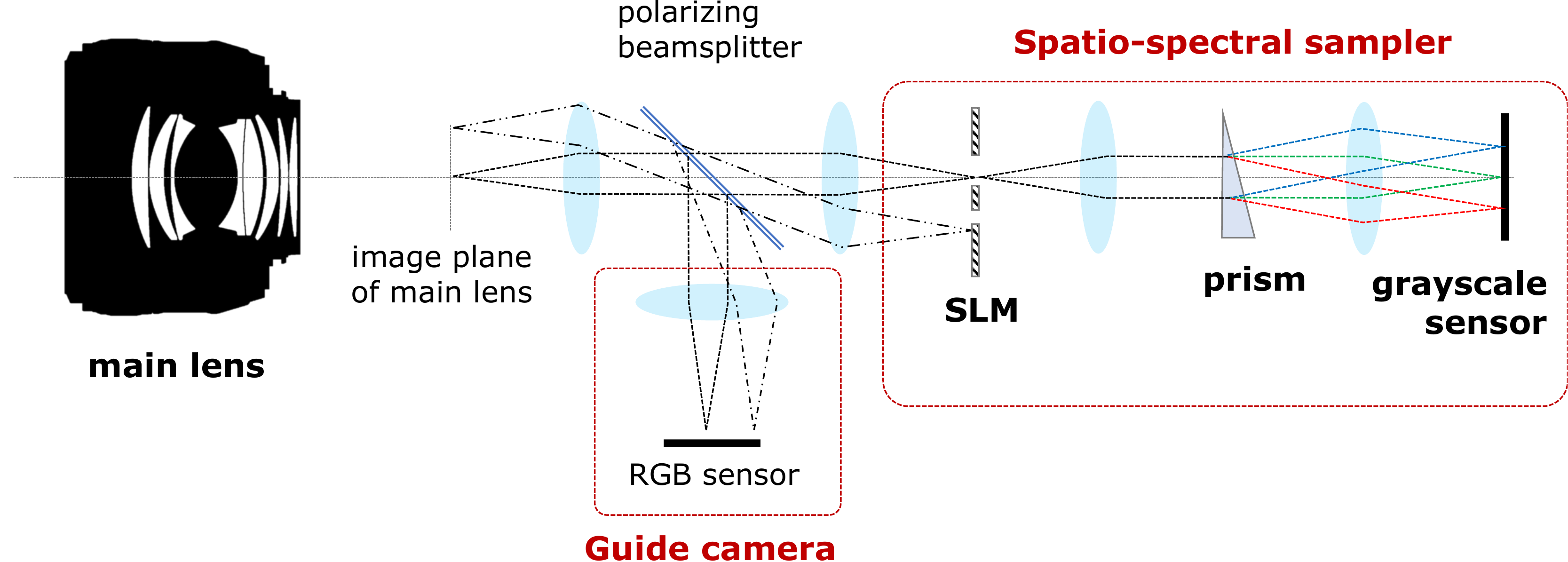}
	\caption{\textbf{Schematic of the proposed setup.} Our optical setup consists of an RGB camera that guides the sampling system, and a spatio-spectral imager that consists of a spatial light modulator, a prism and a grayscale sensor. The guide image is utilized to generate a spatial mask that generates non-overlapping spectral profiles on the grayscale sensor. The guide image is then fused with the sparse spectral measurements to obtain a high spatial and spectral resolution HSI.}
	\label{fig:schematic}
\end{figure}

%
%
We first describe the SASSI system in detail, starting with the optical design, and the key processing steps for reconstruction of snapshot HSIs.
\begin{figure*}[!tt]
	\centering
	\includegraphics[width=\textwidth]{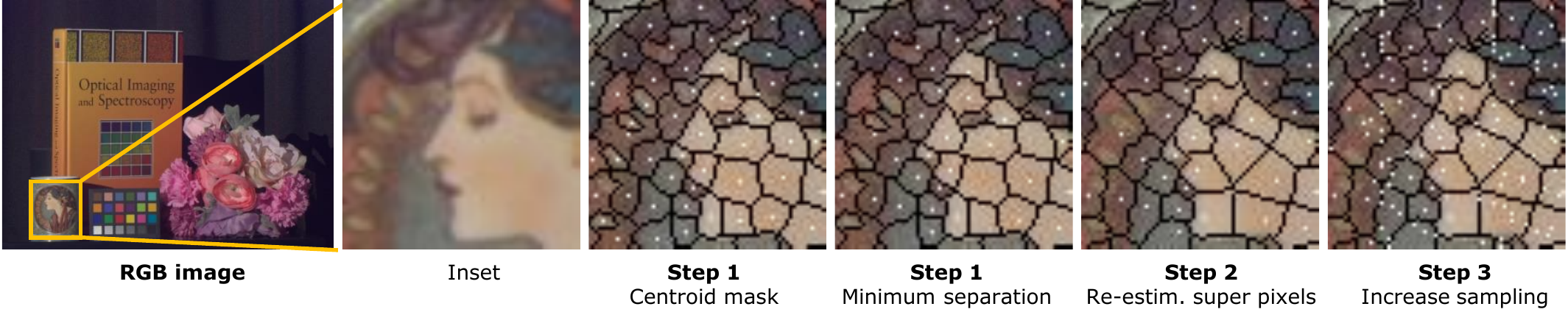}
	\caption{\textbf{Super-pixel sampling strategy.} We follow a computationally simple, three step sampling strategy for estimating the mask. Given a super-pixel segmentation, we first create a mask with centroids of super-pixels as sampling locations. Then, we enforce minimum separation along horizontal direction by moving/removing sampling locations. Next, we re-estimate the super-pixels with the new sampling locations as centroids. Finally, we increase light throughput by creating sampling locations everywhere with minimum separation between neighbors. Such a sampling strategy requires less than $2$ ms on a modern computer, ensuring real time mask generation.}
	\label{fig:sampling_strategy}
\end{figure*}

\subsection{Optical schematic}
Figure \ref{fig:schematic} shows a schematic of our optical setup, which consists of two sub-systems: the guide RGB camera and the spatio-spectral sampler that involves an SLM for spatial sampling, a prism for spectral dispersion, and finally a grayscale camera. 

Given the HSI of the scene, $H(x, y, \lambda)$, we now describe the image formation process at both cameras.
An image of the scene is focused by a main imaging lens.
The image plane is relayed to both the RGB guide camera and the SLM via relay optics and a polarizing beamsplitter.
The guide camera senses a color image of the scene, $I_\text{RGB}(x, y)$, whose intensities relate to $H(x, y, \lambda)$  via the sensor's spectral response curves.
The SLM implements a transmission mask $M(x, y)$ which results in a modified HSI, $\widetilde{H}(x, y, \lambda) = H(x, y, \lambda) M(x, y)$. 
This signal then propagates through the beamsplitter and the prism which generates the following spectrally-spread image,
\begin{align}
	I_\text{spec}(x, y) &= \int_\lambda \widetilde{H}(x - f(\lambda), y, \lambda)\ c(\lambda)\ d\lambda, 
	\label{eq:imform}
\end{align}
where $f(\cdot)$ is a wavelength dependent shift induced by the prism and $c(\lambda)$ is spectral response of the camera.
Note that (\ref{eq:imform}) is the image formation model for the single-disperser CASSI system, if we have a static  attenuation mask $M(x, y)$.
We assume the maximum spatial spread of spectrum to be $N$ pixels.
Note that, with careful design of the mask $M(x, y)$, we can ensure that each pixel in the image $I_\text{spec}$ only measures light from a single spatial pixel of the SLM.
%

Given the image formation model above, the overall sensing pipeline is as follows.
The guide image acquired at time $t-1$ is used to determine the mask at time $t$, $M(x, y, t)$, via super-pixelation and judicious sampling of pixels given the super-pixels.
The mask $M(x, y, t)$ is deployed on the SLM when acquire the  images at time $t$; these images, along with the mask, are fused to obtain the hyperspectral image at time $t$. We describe each step in detail next.

\subsection{Super-pixel generation}
%
%
Given the guide image, we perform super-pixelation using off-the-shelf techniques. 
We used Simple Linear Iterative Clustering (SLIC) \cite{achanta2012slic} algorithm as it is computationally lightweight \cite{fastslic2020}.
%
%
The SLIC algorithm essentially has two parameters, namely the number of super-pixels, $Q$ and the \emph{compactness}, $C$.
Compactness controls the regularity of super-pixels; a larger value of $C$ promotes more regularly shaped super-pixels, while smaller value ensures that the super-pixel boundaries adhere to edges in the image.
Such a simple and effective parametrization is favorable to our mask generation scheme, and hence very crucial.
The size and shape (compactness) of super-pixels depends on number of pixels over which spectrum is spread at each spatial point and hence can be effectively set.
%

\subsection{Adaptive spatial sampling}
Given the super-pixel image, we now wish to efficiently generate a scene-adaptive sampling mask that provides maximum amount of information about the HSI, without spectral multiplexing.
%
%
The generated mask aims to satisfy  two key requirements:
\begin{enumerate}
	\item Each super-pixel is sampled at least once, and
	\item The sampling locations are horizontally separated by $N$ pixels to ensure no  overlap between the spectral smears arising from two different spatial locations.
\end{enumerate}
In addition to these two requirements, to increase light efficiency, we would like to allow light to enter from as many SLM pixels as possible.
While an optimal mask generation is a computationally expensive problem, we found the following multi-step approach worked very efficiently.
%

{{\flushleft \textbf{Step 1 -- Initial centroid-based mask.}}} We start by setting the sampling locations to superpixel centroids, $\{(x_c^k, y_c^k)\}_{k=1}^Q$.
This generates a mask $M_1(x, y)$ that has $Q$ openings in total, which may be small but ensures that all super-pixels are sampled.
%
%
To enforce the second constraint, we then modify the initial mask by moving/removing sampling locations till every location is separated by at least $N$ pixels.

{{\flushleft \textbf{Step 2 -- Resegmentation of super-pixels.}}}
The resultant mask after step 1 satisfies the second constraint but may violate the first constraint of one sample per superpixel -- especially if the super-pixels are small.
In order to satisfy the first constraint, we use the updated sampling locations from first step as the centroids and reestimate the super-pixels. In case of SLIC super-pixels, this involves running a simple local K-means clustering which is fast and efficient.

{{\flushleft \textbf{Step 3 -- Increasing light transmission.}}}
Given a spatial resolution of $H\times W$ and a spectral spread of $N$ pixels, the mask can have a maximum of $Q_\text{max} = \frac{HW}{N}$ sampling locations.
%
We followed a greedy approach to achieve this.
Starting from left side of the mask, we open every location that is not sampled in a $2N-1$ horizontal neighborhood.
This is then repeated for all rows of the image till no more pixels can be opened.
Figure \ref{fig:sampling_strategy} visualizes each step of the sampling strategy. Given a super-pixel segmentation, the whole process takes less than $2$ ms on a mid-range PC.

\subsection{Reconstruction techniques}
We propose two approaches to estimate HSIs.

{{\flushleft \textbf{Local rank-1 approximation.}}}
We follow a ``rank-1" approximation for each super-pixels, where we assume that spectrum at each spatial location is a scaled version of the sampled spectra within its super-pixel neighborhood.
%
%
Let $I_\text{gray}(x, y)$ be the grayscale image of the scene obtained from the guide camera.
We illustrate the reconstruction details for one super-pixel; the method is the same for all other super-pixels.
Let $\{(x_{p}, y_{ p})\}_p$ be sampled locations within the $l^\text{th}$ super-pixel, and let $S_p(\lambda) = H(x_{p}, y_{p}, \lambda)$ the sampled spectrum.
The spectrum at an unsampled location $(x_{u}, y_{u})$  within this super-pixel is then computed as,
\begin{align}
	\widehat{H}(x_{u}, y_{u}, \lambda) = \frac{I_\text{gray}(x_{u}, y_{u})}{\sum_{p=1}^{N_l} I_\text{gray}(x_{p}, y_{p})} \sum_{p=1}^{N_l} S_{p} (\lambda)
\end{align}
Such a reconstruction strategy is computationally inexpensive, and can be easily implemented on GPUs, making it an appealing approach for real-time reconstruction.

{{\flushleft \textbf{Learned guided filtering.}}}
When capturing measurements at video rate, the SNR is expected to be low, which gives rise to artifacts around superpixel boundaries.
To overcome this, we rely on a data-driven reconstruction for high quality reconstruction even under severely noisy measurements.
The approach is similar in spirit to the well-known guided reconstruction techniques such as guided filtering \cite{he2010guided}, and more recently, a neural-network variant of the same \cite{wu2018fast}.

Our neural network based reconstruction has two distinct components: first, a guided filter with learnable filters, and second, a simple 2-layer neural network. We describe each below.

\textit{Step 1 ---- Guided filter}. The first step estimates the spectral image as an affine scaling of the guide, i.e.\, $I^k_\lambda(x, y) \approx  \alpha_k I^k_\text{guide} +\beta_k$ for the $k^\text{th}$ image patch.
%
%
The values of $\alpha_k$ and $\beta_k$ are estimated on the fly by solving an optimization problem of the form,
\begin{align}
	\min_{\alpha_k, \beta_k} \|F\odot M^k\odot \left( I^k_\text{input} - \alpha_kI^k_\text{guide} -\beta_k\right)\|^2,
\end{align}
where $M^k$ is the spatial mask for the $k^\text{th}$ patch, and $F$ is a box function in traditional guided filtering.
We observe that the box function is not sufficient when working with sampled data, and hence we make the parameters of $F$ trainable as part of the neural network.

\textit{Step 2 ---- Refinement.} We then use the outputs of the guided filtering layer as inputs to a simple two layer neural network, which acts as a refinement layer to remove any artifacts.
Other architectures based on guided filtering are possible, but we found the simple three-layer architecture sufficed for our purposes.

\begin{figure}[!tt]
	\centering
	\includegraphics[width=\columnwidth]{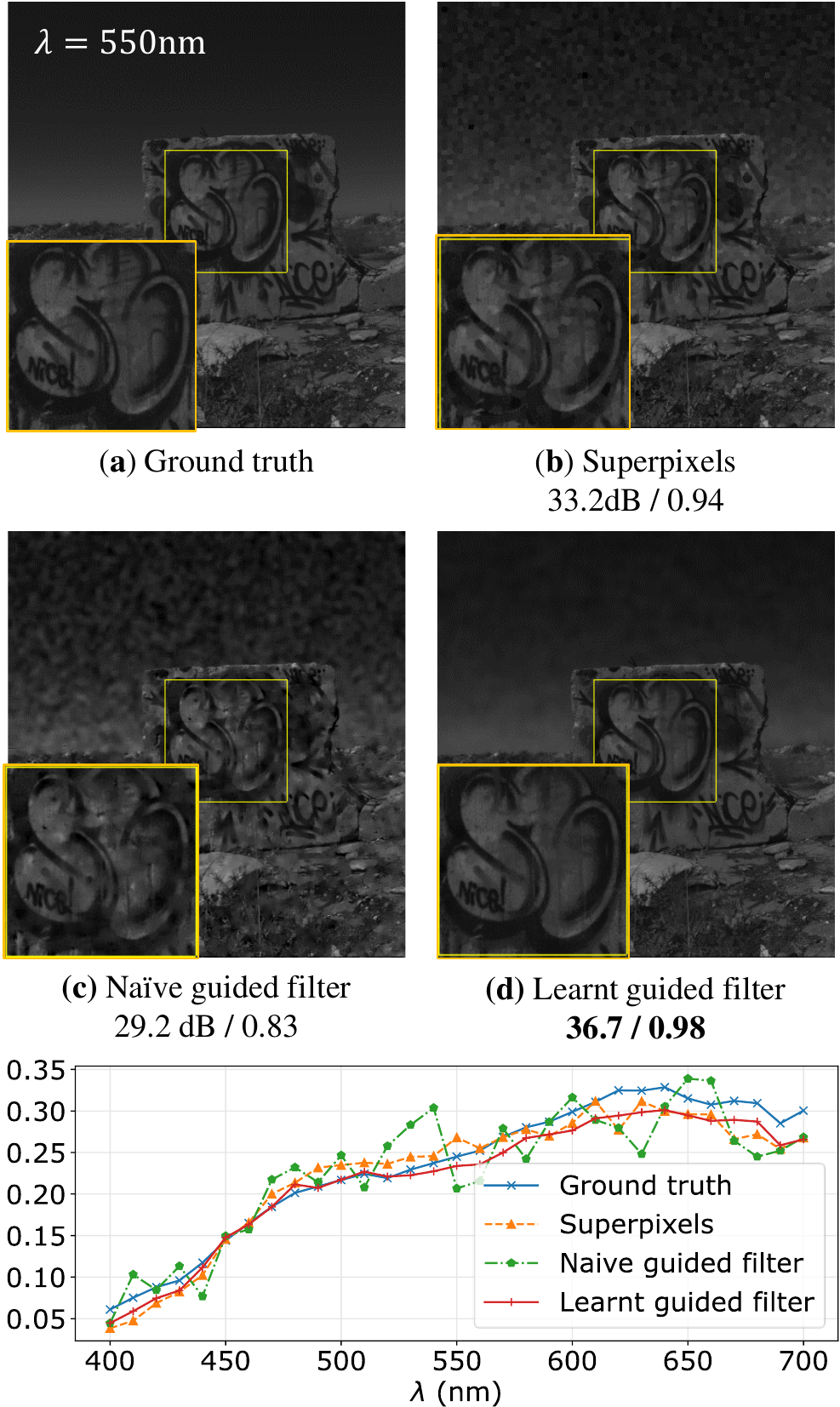}
	\caption{\textbf{Data-driven reconstruction.} We rely on a modified guided filter to reconstruct HSIs from RGB and spectral measurements. Compared to superpixel-based reconstruction or naive guided filter, our learning-based is more robust to noise, and hence better suited to low SNR conditions.}
	\label{fig:sp_gd_nn}
\end{figure}

\begin{figure}[!tt]
	\centering
	\includegraphics[width=\columnwidth]{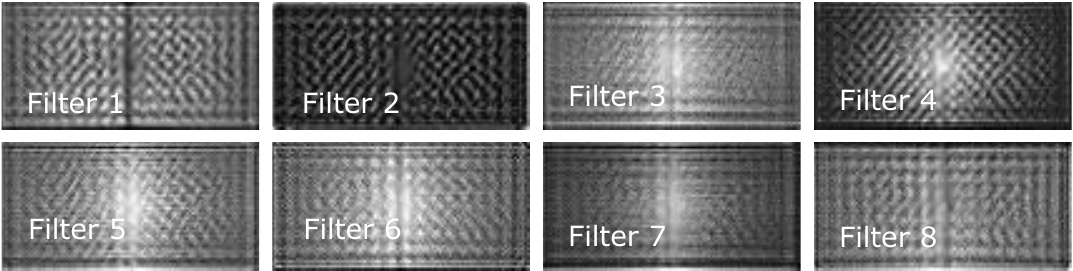}
	\caption{\textbf{Learned guided filters.} We optimized for a total of $8$ guided filters of $51\times103$ dimension. While naive guided filters are uniform box filters, we note that learned ones have a shape that promotes higher reconstruction accuracy for our sparsely sampled HSIs.}
	\label{fig:sim_filters}
\end{figure}

Figure \ref{fig:sp_gd_nn} illustrates the benefits derived from using learned stack of filters.
Under very low photon levels, our rank-1 approximation has superpixel artifacts.
Naive guided filtering partly alleviates this but causes hazy reconstructions.
In contrast, our guided filtering approach gives accurate reconstruction accuracy.
Figure \ref{fig:sim_filters} visualizes the learned guided filters on the KAIST~\cite{deepcassi2017} dataset.
We observe that the shape of the filters is significantly different from a box filter, and is better to our setting.
%
%


%
%
%

\begin{figure}[!tt]
	\centering
	\includegraphics[width=\columnwidth]{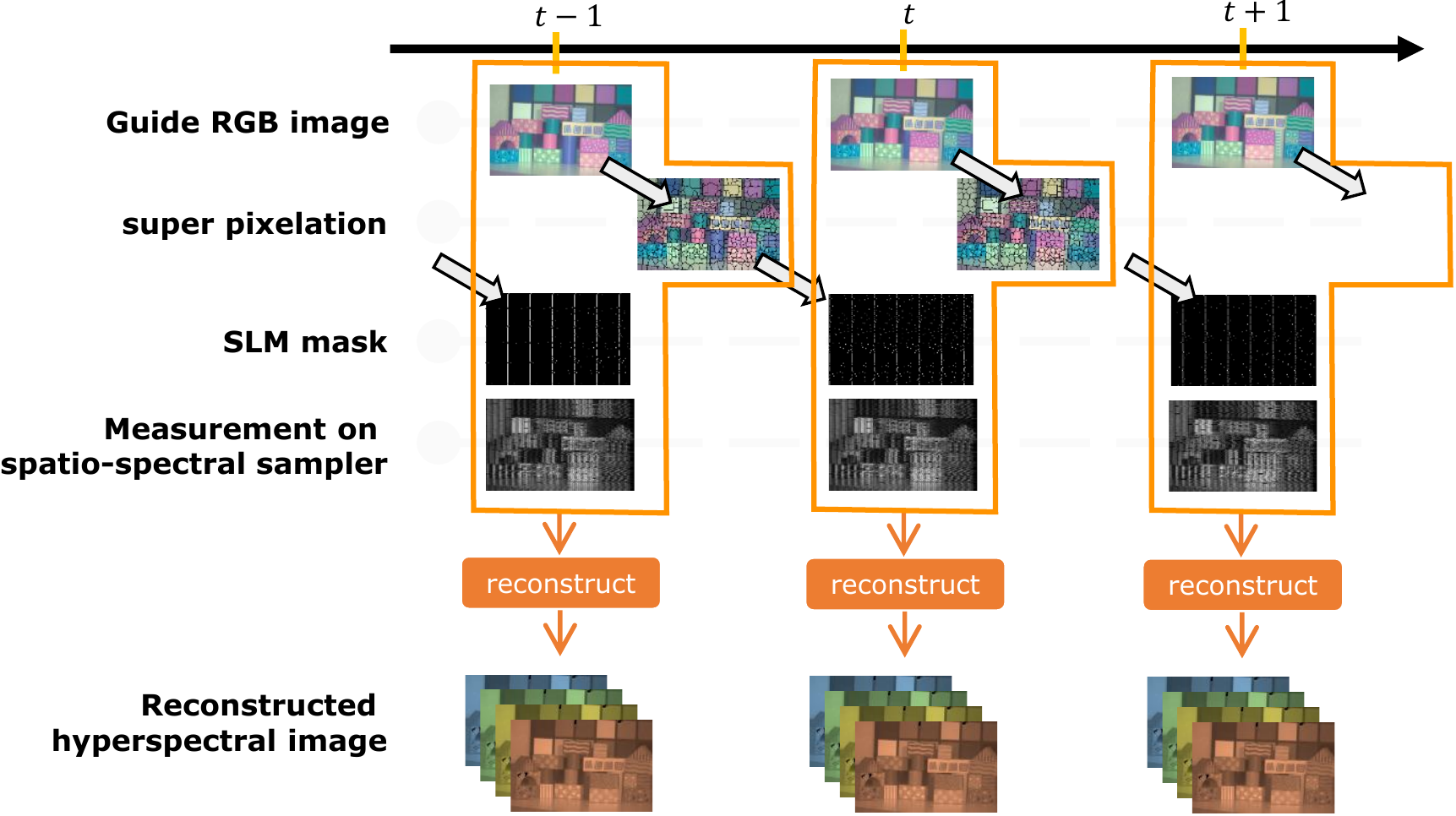}
	\caption{\textbf{System pipeline.} Our method avoids temporal registration  as it relies only on the information captured at time instance $t$. Specifically, the guide image at $t-1$ is used to create a mask, which is used to capture another guide image and the spatial-spectral image at time $t$. Instead of fusing guide image from $t-1$, we fuse images only from time instance $t$ which ensures that there are no motion artifacts.}
	\label{fig:video_sampling}
\end{figure}

\subsection{Reconstruction approach for video sequence}
%
Since our approach requires fusion of two images, it is important they capture measurements of the \emph{same} scene.
When capturing a video sequence, our optical system generates color images $I^t_{RGB}(x, y)$ which generates masks $M^t(x, y)$ at  time instance $t$. 
This mask is used to measure the RGB and spectral images at time instance $t+1$, which is likely different from the scene at $t$ due to motion.
Instead of fusing $I^t_\text{RGB}(x, y)$ and $I_\text{spec}^{t+1}(x, y)$, which may lead to erroneous reconstruction, we combine $I^{t+1}_\text{RGB}(x, y)$ and $I_\text{spec}^{t+1}(x, y)$ --- which ensures that both images are in lock-step.
We note here that step 2 in the mask generation process, where the super-pixels are re-estimated is an important contributor to accurately fusing the two images.
By generating mask from time instance $t$ and using it to estimate super-pixels in time instance $t+1$, we avoid temporal registration which is hard to perform for an arbitrary scene.
The timing of the system pipeline is illustrated in Figure \ref{fig:video_sampling}.


\section{Experiments} \label{section:experiments}
We first validate SASSI with simulations on a commonly available datasets, and then demonstrate results with the optical setup we built in our lab. 
%

\subsection{Simulations}
We first simulated our approach to understand the advantages of adaptive sensing, impact of noise, and choice of number of superpixels.

\bpara{Datasets.} We simulated our approach on standard HSI datasets including CAVE dataset \cite{yasuma2010generalized}, KAIST dataset \cite{deepcassi2017}, Harvard Dataset \cite{chakrabarti2011statistics}, and ICVL dataset \cite{arad_and_ben_shahar_2016_ECCV}, each with $31$ spectral bands from $400-700$nm.
%
%
For video rate results, we used data from \cite{mian2012hyperspectral} which comprised of 31 video frames, with each frame consisting of $33$, $480\times752$ images sampled from $400 - 720$nm.

\bpara{Modeling noise.} 
%
We modeled our sensing camera to have a readout noise standard deviation of $5 e^-$, a value typical to scientific cameras.
We assumed various light levels, starting from $100 e^-$ per pixel to $10,000 e^-$  per pixels with the aim of showing performance variations with noise levels.

\bpara{Training details.} 
%
Our neural network consisted of $8$ guided filter layers of $51\times103$ dimension each, followed by two convolutional layers along with ReLU non-linearity, each with 32 layers of $3\times3$ filters.
%
%
Further details about the training process can be found in the appendix.

\subsubsection{Number of superpixels}
%
%
The number of superpixels is upper bounded by $\frac{HW}{N}$; however, the specific number of super pixels that maximizes reconstruction accuracy is depends on spatial complexity of the scene, as well as the noise levels.
Intuitively, smaller super pixels ensure that highly textured scenes are well sampled.
However, each super pixel will then get fewer samples, which is not desirable in low light conditions.

\begin{figure}[!tt]
	\centering
	\begin{subfigure}[t]{\columnwidth}
		\centering
		\includegraphics[width=\columnwidth]{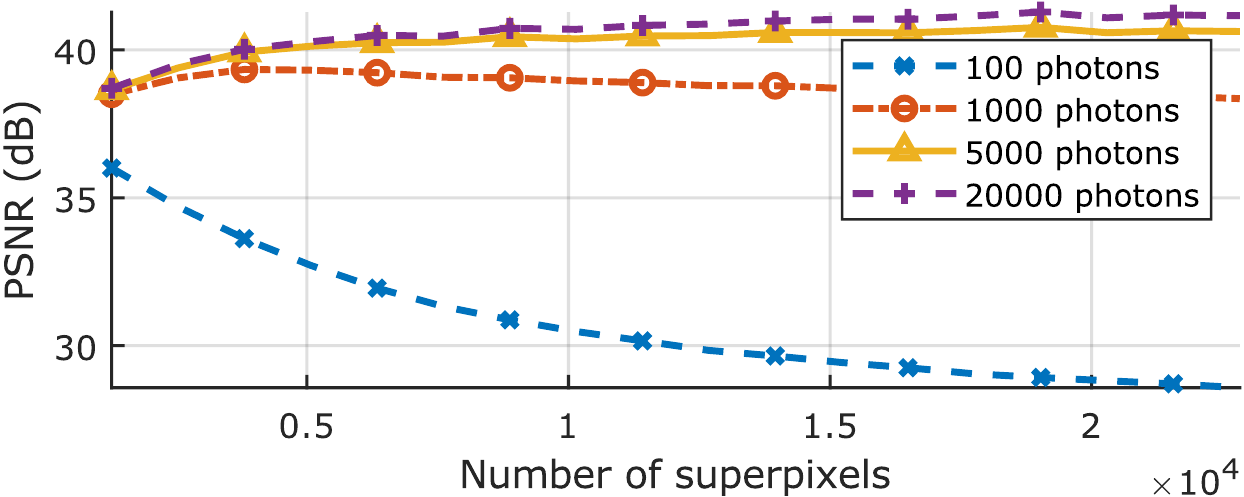}	
		\caption{Performance with number of superpixels}		
		\label{fig:snr_vs_nsup}	
	\end{subfigure}
	\begin{subfigure}[t]{0.48\columnwidth}
		\centering
		\includegraphics[width=\textwidth]{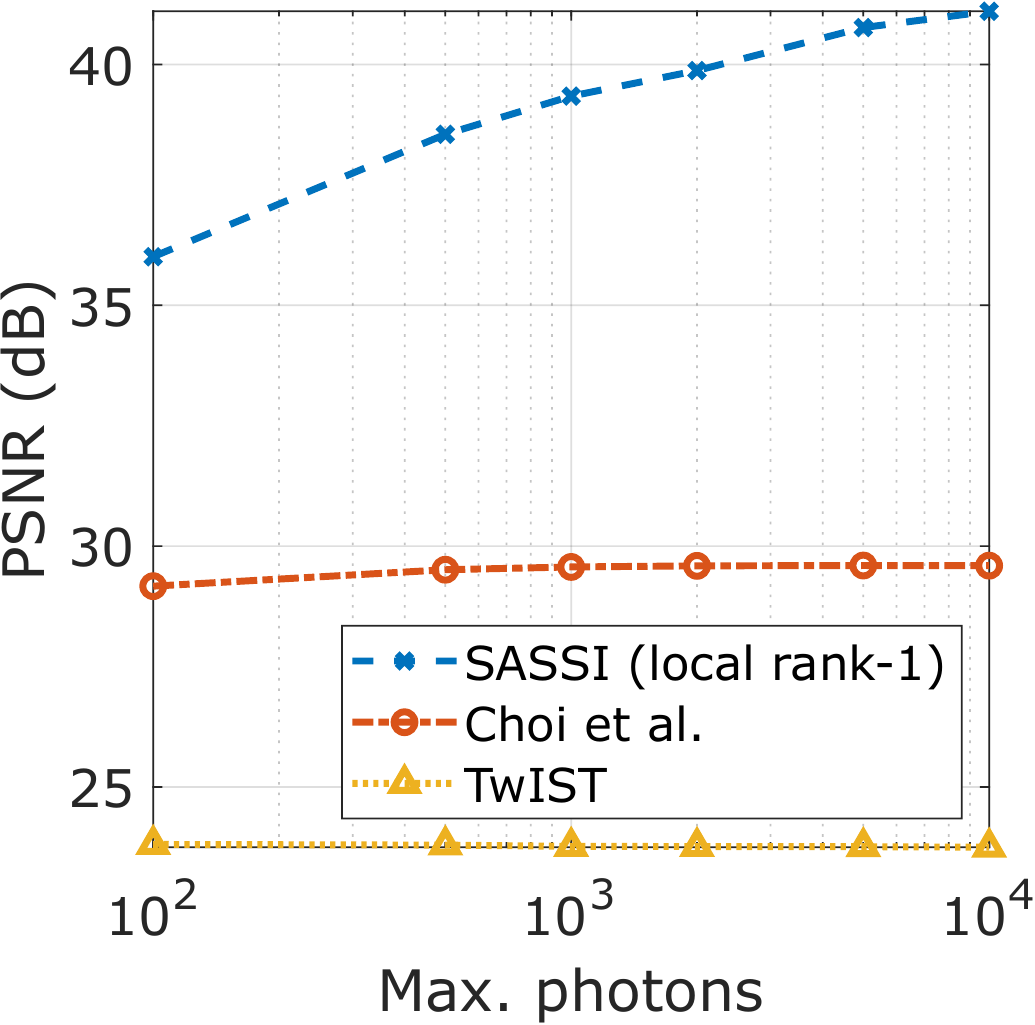}
		\caption{Average to high intensity}
	\end{subfigure}
	\begin{subfigure}[t]{0.48\columnwidth}
		\centering
		\includegraphics[width=\textwidth]{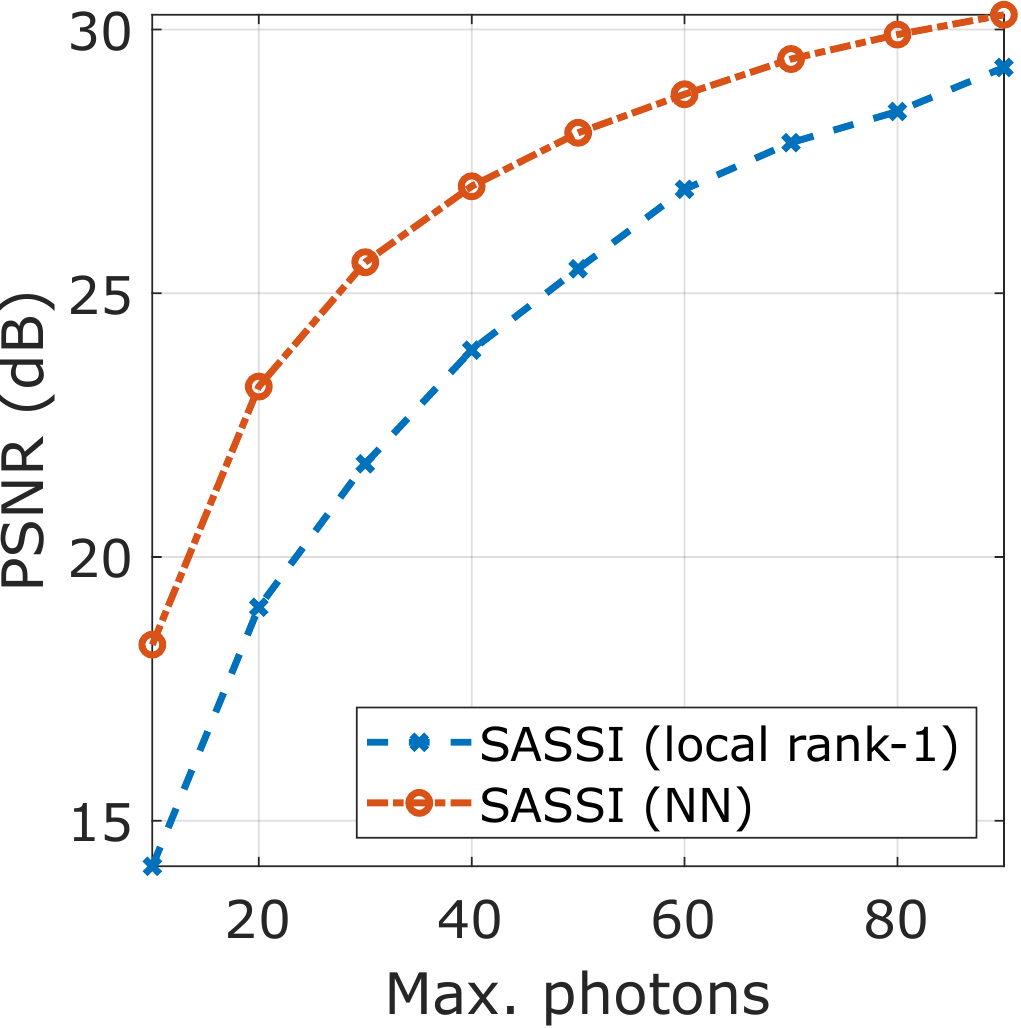}
		\caption{Low light levels}
	\end{subfigure}
	\caption{\textbf{Performance with number of superpixels and noise.} (a) As noise increases, a small number of superpixels is advantageous, as the spectra within a superpixel get averaged. (b) Our rank-1 reconstruction performs well against prior work under average to high light intensity. (c) Tt lower light levels, our guided filtering approach outperforms the rank-1 approach.}
	\label{fig:snr_vs_noise}
\end{figure}

We empirically evaluated the effect of number of super pixels under varying noise conditions on sample HSIs.
%
Figure \ref{fig:snr_vs_nsup} showed accuracy as a  function of number of superpixels under varying light conditions.
At low light, larger superpixels are advantageous, while at higher light levels, smaller superpixels lead to accurate results.
%
%
%
In our real experiments, we found that $Q = \frac{1}{4}\frac{HW}{N}$ gave the best results for short exposure times of $50$ ms or lower.


\begin{figure*}[!tt]
	\centering
	\includegraphics[width=\textwidth]{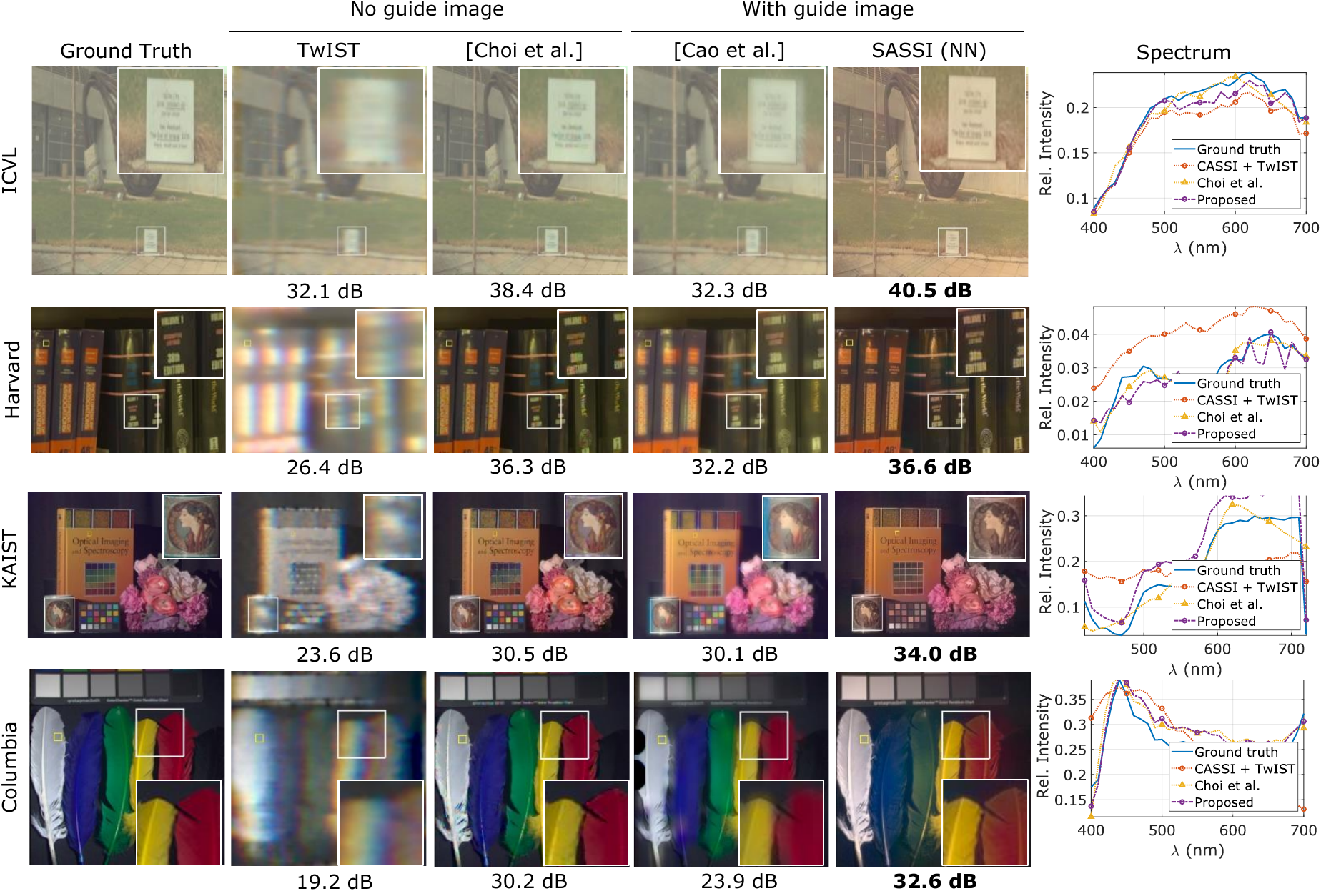}
	\caption{\textbf{Comparisons with CASSI techniques.} We compared reconstruction approaches which do not use a guide image \cite{deepcassi2017, bioucas2007new}, and ones that use \cite{cao2011high} which includes our approach. Across the board, our method shows higher accuracy.}
	\label{fig:simulations_master}
\end{figure*}
\subsubsection{Comparisons to prior art}
We simulated densely-sampled mask based reconstruction approaches with traditional total variation (TV) penalized reconstruction \cite{bioucas2007new} and a more recent neural network based approach \cite{deepcassi2017}, denoted by TwIST and [Choi et al.], respectively.
Figure \ref{fig:snr_vs_noise} compares our method against \cite{deepcassi2017} and reconstruction with TwIST~\cite{bioucas2007new} across varying light levels, and Fig. \ref{fig:simulations_master} visualizes the reconstructions. 
These methods do not utilize an extra guide image.
We also compared against guide image based reconstruction approach by [Cao et al.]\ \cite{cao2011high} where the HSI is uniformly and sparsely sampled and reconstruction is done through a modified bilateral filtering.
%
%
Among methods that do not utilize a guide image,  \cite{deepcassi2017} had superior performance in terms of visual quality as well as reconstruction SNR.
When using a guide image, our approach outperforms other approaches by more than $2$dB.
Our approach is also several orders of magnitude faster than other approaches. For a $256\times256\times31$ dimensional HSI, \cite{bioucas2007new} approach took 10s on CPU, \cite{deepcassi2017} took $1.4$ minutes on GPU, \cite{cao2011high} took $3.1$ minutes on CPU, while our approach took less than $5$ seconds on the GPU.

%

\subsection{Experiments with a lab prototype}
%
%
We first provide details of our optical setup and then demonstrate results in various settings.

\subsubsection{Optical setup}
\begin{figure}[!tt]
	\centering
	\includegraphics[width=\columnwidth]{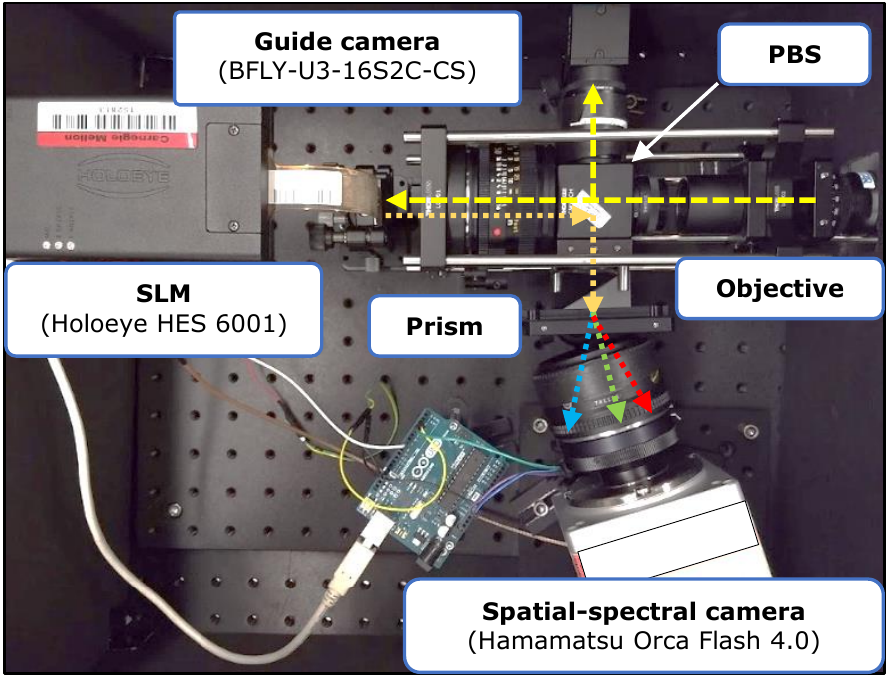}
	\caption{\textbf{Lab prototype.} The image above shows a photograph of our lab setup with key components marked. We also showed an overlay of ray tracing for easy understanding.}
	\label{fig:lab_setup}
\end{figure}

Figure \ref{fig:lab_setup} shows an image of the optical setup.
We used a FLIR Blackfly BFLY-U3-16S2C-CS color camera as guide camera, and Hamamatsu Orca Flash 4.0 as spatial-spectral camera, Holoeye HES 6001 LCoS display as SLM operating at 60fps, and a 
%
%
a $30^\circ$ prism for dispersion.
Additional details can be found in the appendix.

{{\flushleft \textbf{System properties.}}} Our setup was optimized to image from $400-700$ nm, with spectrum spread over $68$ pixels on the spatial spectral camera.
Due to non-linear dispersion of the prism, we obtain a spectral resolution of $2$ nm at $400$nm and $10$ nm at $700$ nm.
All the HSIs captured in the upcoming sections were captured at a spatial resolution of $698\times931$ pixels.
Detailed calibration steps required to scan HSIs are provided in the appendix.

\begin{figure}[!tt]
	\centering
	\begin{subfigure}[t]{0.3\columnwidth}
		\centering
		\includegraphics[width=\columnwidth]{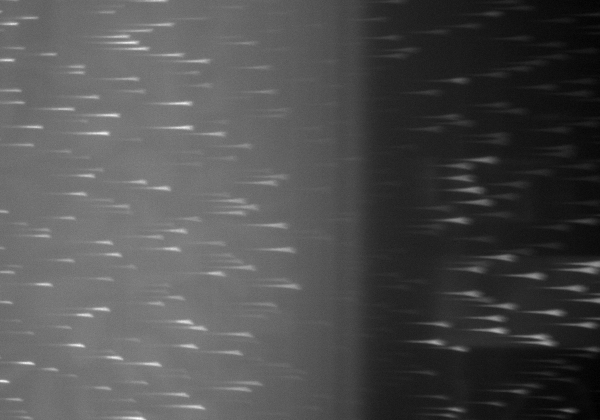}
		\caption{Raw image}
	\end{subfigure}
	\hspace{0.1em}
	\begin{subfigure}[t]{0.3\columnwidth}
		\centering
		\includegraphics[width=\columnwidth]{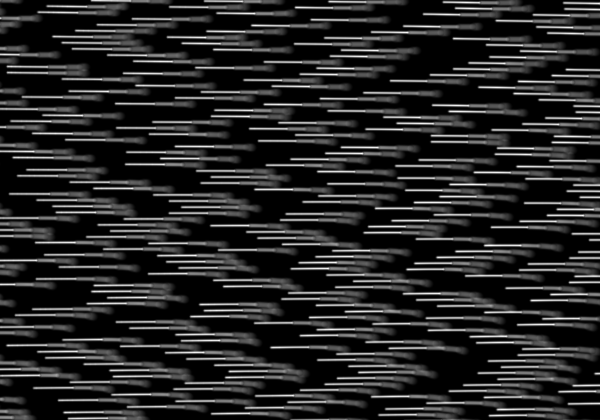}
		\caption{Predicted streaks}
	\end{subfigure}
	\hspace{0.1em}
	\begin{subfigure}[t]{0.3\columnwidth}
		\centering
		\includegraphics[width=\columnwidth]{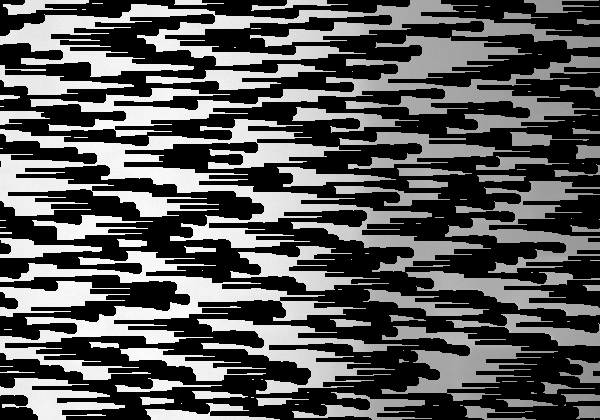}
		\caption{Image without streaks}
	\end{subfigure}
	\\
	\begin{subfigure}[t]{0.3\columnwidth}
		\centering
		\includegraphics[width=\columnwidth]{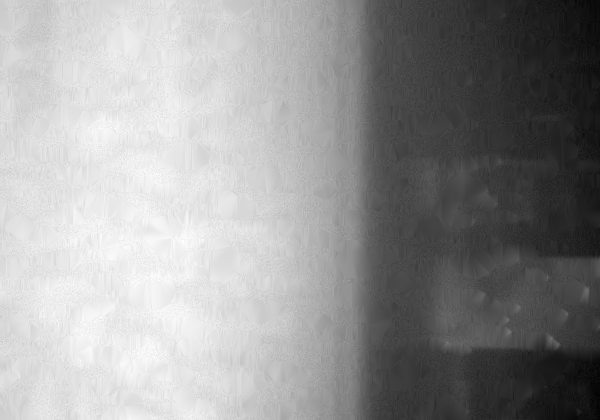}
		\caption{Computed offset}
	\end{subfigure}
	\hspace{0.1em}
	\begin{subfigure}[t]{0.3\columnwidth}
		\centering
		\includegraphics[width=\columnwidth]{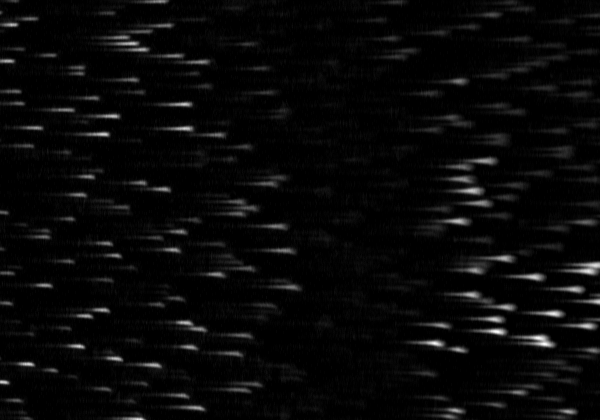}
		\caption{Clean image}
	\end{subfigure}
	\hspace{0.1em}
	\begin{subfigure}[t]{0.3\columnwidth}
		\centering
		\includegraphics[width=\columnwidth]{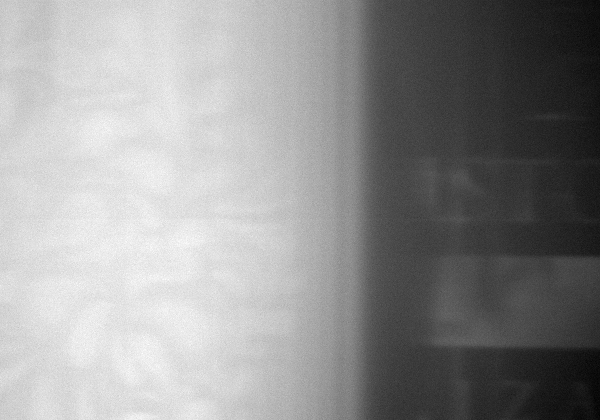}
		\caption{Image with all-zeros mask}
	\end{subfigure}
	\caption{\textbf{Removing background offset.} Since the SLM has finite contrast ratio, the resultant mask has non-zero values everywhere, leading to a background offset, shown in (a). We rely on (b) calibration information, along with interpolation to predict the offset image in (d), leading to a cleaned image in (e). Notice that the offset image is close to an image captured with all-zeros spatial mask.}
	\label{fig:offset}
\end{figure}

\bpara{Removing offset in spectral image.}
%
Due to finite contrast ratio of SLMs, the spatial mask can be effectively written as $M_\text{real} = (1-\epsilon)M_\text{ideal} + \epsilon$, where $\epsilon$ is the non-zero throughput when the SLM pixel is set to $0$. 
This leads to a large background offset in the measured spatial-spectral image, as illustrated in Fig. \ref{fig:offset}(a).
This can be compensated by capturing an image with an all-zero mask, like Fig. \ref{fig:offset}(f); however, this is not feasible when operating with time-varying scenes.
%
%
Instead we observe that the offset image is smooth and estimate it from the spatiao-spectral image, by first identifying the removing the spectral streaks, and interpolating the missing pixels using cubic interpolation.
This process is visually illustrated in Figure \ref{fig:offset}.

\bpara{Nyquist scanning.} 
%
To obtain a full scan of scene's HSI, we implement a parallelized version of raster scan where a pinhole array is displayed on the SLM and scanned. The horizontal spacing of the pinhole array is kept at 100 pixels, which is greater than the dispersion by the prism, and the verticle spacing is at 5 pixels.
%
%
We then scanned a total of $500$ images, with $100$ horizontal shifts and $5$ vertical shifts.
Figure \ref{fig:sampling_compare} shows an example of the mask that is used as well as the corresponding measurement in this sampling process.



\subsubsection{Dataset details}
\begin{figure}[!tt]
	\centering
	\includegraphics[width=0.475\textwidth]{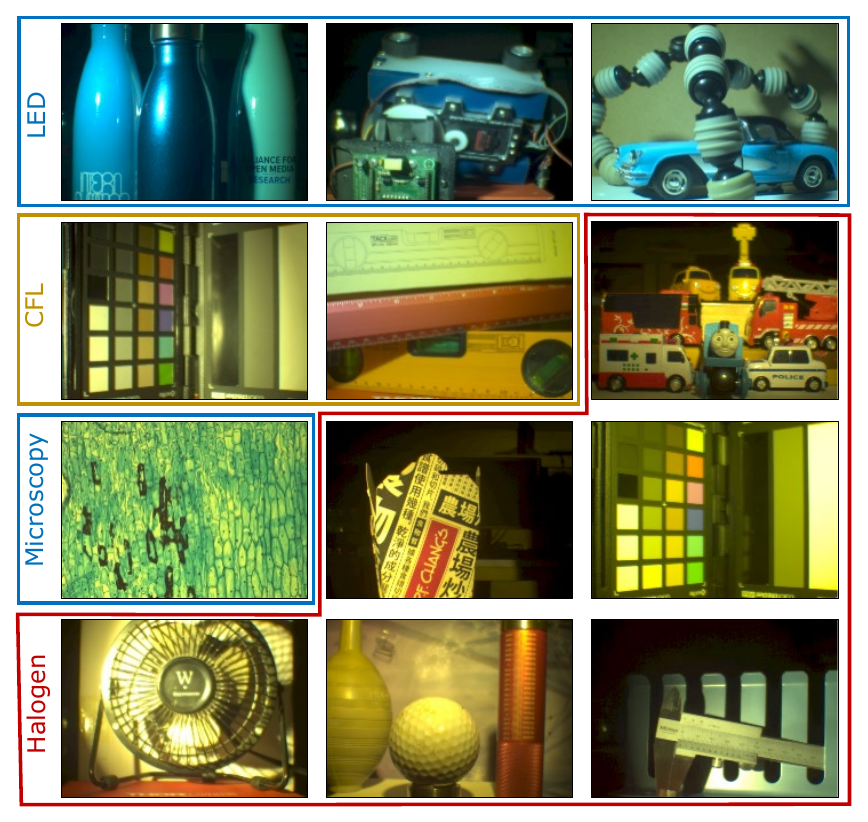}
	\caption{\textbf{Training data for learning guided filters.} We collected more than $50$ hyperspectral images with various objects and illuminants. We then used 18 for training the neural network. None of the training images were used in the testing phase.}
	\label{fig:training_data}
\end{figure}

We collected $55$ hyperspectral images of various objects, under varying illumination conditions.
A snapshot of the training data is shown in Fig. \ref{fig:training_data}.
Each capture included a full scan of the scene, along with a guide RGB image, and the corresponding super-pixel sampled image.
We retrain the guided filter on this dataset for finetuning it to the specifics of the prototype.
%
Further details about the training parameters can be found in the appendix.
%
%

\subsubsection{Results}
We demonstrate results with our setup on static and dynamic scenes, and estimation of spectral reflectance. Unless specified, each measurement involved a single image captured by the guide and CASSI cameras.

\begin{figure}[!tt]
	\centering
	\begin{subfigure}[c]{\columnwidth}
		\includegraphics[width=\columnwidth]{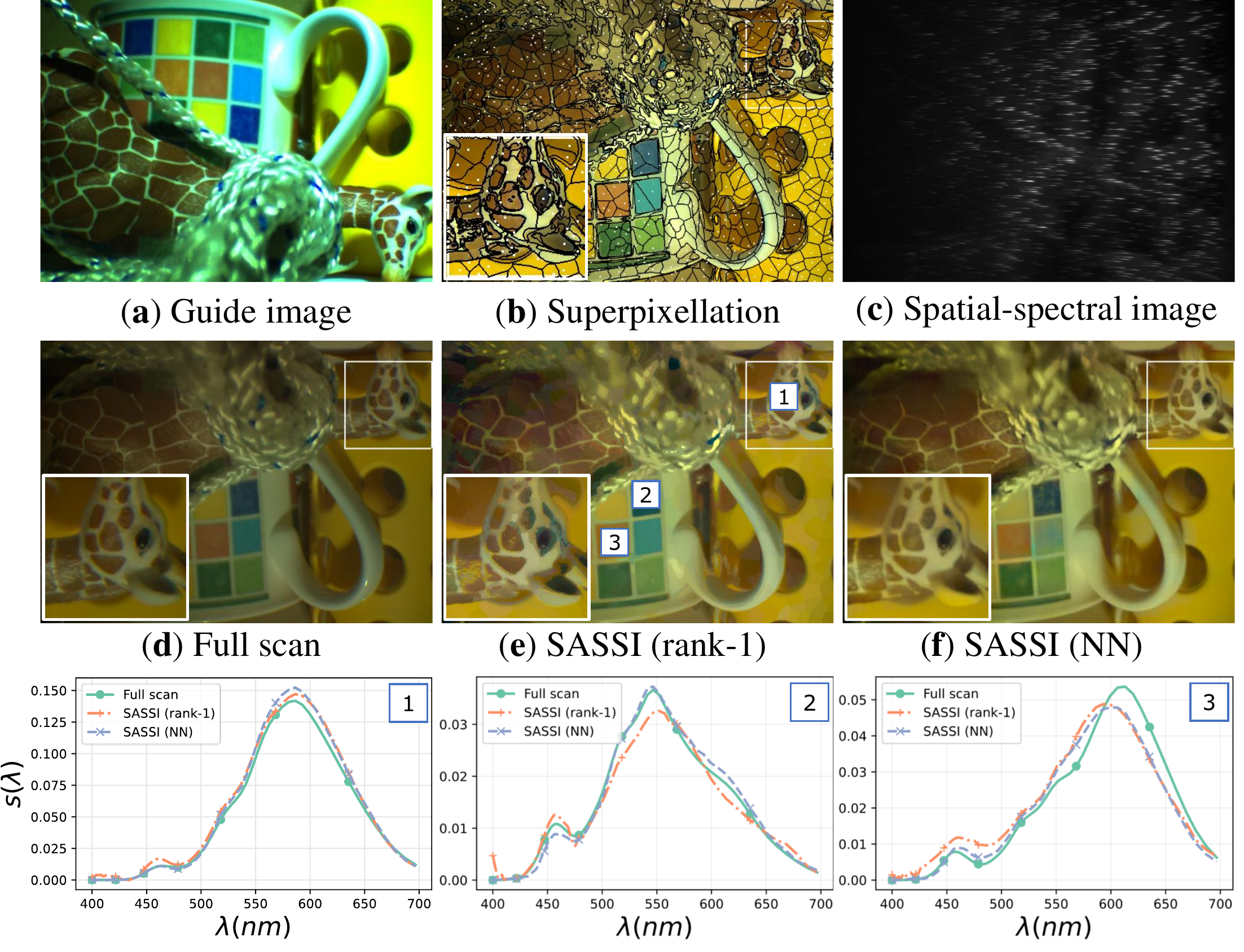}	
	\end{subfigure}
	\begin{subfigure}[c]{\columnwidth}
		\centering
		\begin{tabular}{c}
			\animategraphics[width=\columnwidth,loop]{5}{figures/giraffe/stack/im}{21}{62}
		\end{tabular}
	\end{subfigure}
	\caption{\textbf{Comparison of rank-1 and NN reconstructions.} Reconstruction results on a scene containing color full object. Shown are  (a) guide image, which is used to generate (b) super pixels and sampling pattern (white dots), and subsequently (c) the spatio-spectral measurements. In (d-e) and the plots, we compare reconstructions with a rank-1 and NN approaches along with full pushbroom scan. The last row is a spectral sweep video for the full  scan and the NN reconstruction. The video is best viewed in Adobe Reader.}
	\label{fig:real_sampling}
\end{figure}

\begin{figure*}[!tt]
	\centering
	\includegraphics[width=\textwidth]{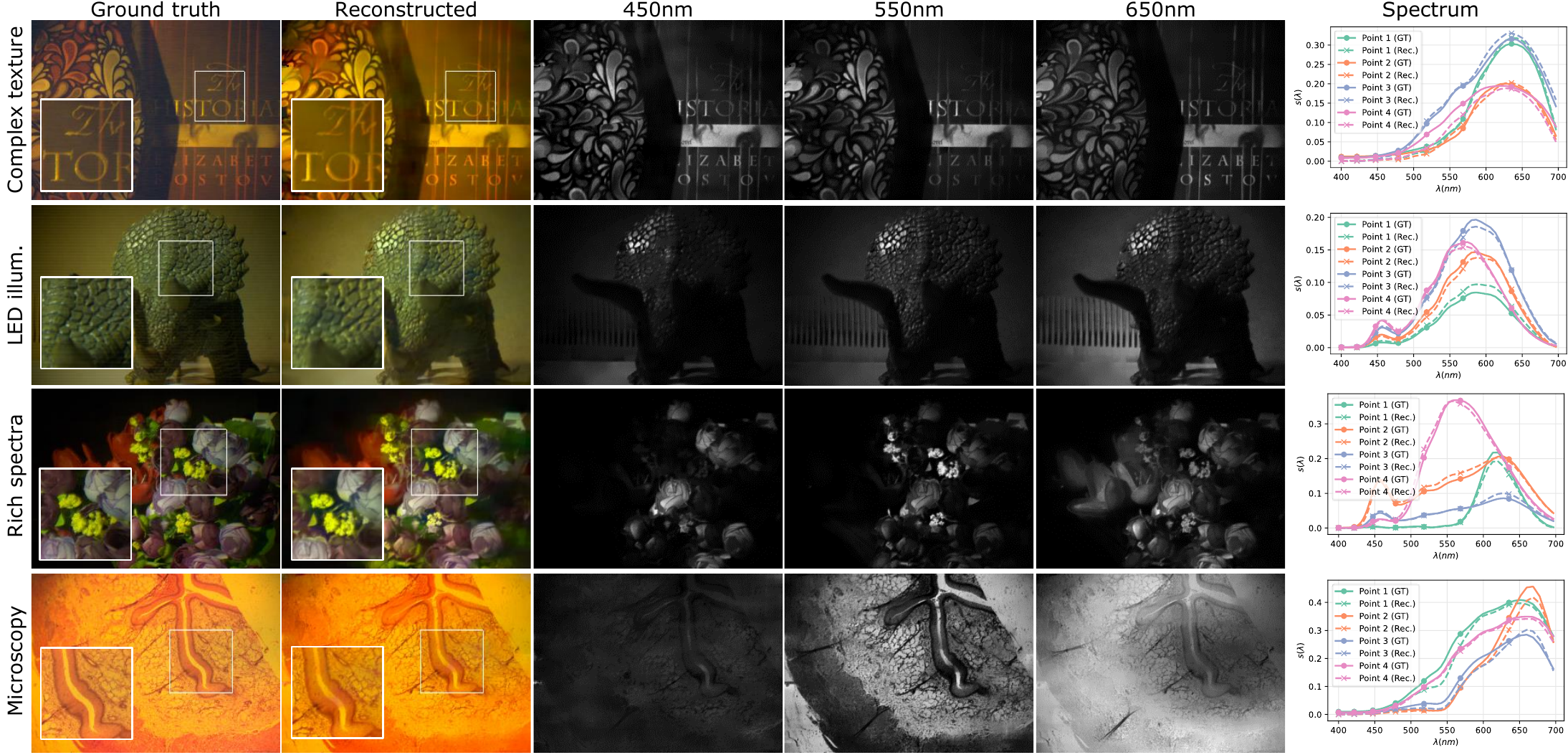}
	\caption{\textbf{Real experiments.} The figure showcases results across a variety of scenes with varying scene, spectrum, and illumination complexity. Across the board, our sampling and reconstruction approach produces accurate results with reconstruction accuracy exceeding $32$dB. }
	\label{fig:real_master}
\end{figure*}

\bpara{Comparison of reconstruction approaches.}
Figure \ref{fig:real_sampling} shows guide image, super pixelation, captured spatial-spectral image and an RGB visualization of reconstruction with rank-1 and  guided filters approach.
Images were captured at an exposure rate of $33$ms.
The second row shows spectra at the marked location, and the third row visually compares ground truth (right) and guided filters reconstruction (left).
%
%
%
The accuracy of rank-1 reconstruction was $41.5$dB with an SSIM of $0.98$, whereas the guided filters approach had an accuracy of $42.6$dB with SSIM of $0.99$.
We observed the performance to be similar across all our real experiments.
This is expected, as the guided filters approach is particularly well suited for low light levels.
%
%
%
The last row shows an animation of images for different wavelengths for both ground truth (left) and our method (right). 

\bpara{Static scenes.}
Figure \ref{fig:real_master} visualizes reconstruction across a variety of geometries, spectral profiles, and illumination conditions.
None of the scenes shown in the figure overlap with the training data for neural networks.
We observe the importance of per-band reconstruction in the third row -- since we only learn on images, and not on the full HSI, we were able to reconstruct complex spectra such as a scene illuminated with LED lamp.
We also tested our system on microscopic specimens, specifically a small tissue of a dog's esophagus.
Notice the difference in intensities of the cellular wall across various wavelengths.
%
%
Across the board, we observed that our sampling and reconstruction approach gave high quality results, with PSNR exceeding $30$dB compared to a full scan of the scene.

\bpara{Dynamic scenes.}
%
SASSI is most impactful when capturing video rate HSIs of scenes with complex geometry and a rich gamut of spectral profiles.
Our setup was capable of capturing HSIs at a rate of $18$ frames per second.
This capture rate was primarily limited by the vagaries associated with our SLM, and not an upper limit on the method itself.
%
%
%
%
Timing analysis of each step of our algorithm is listed in Tab. \ref{tab:timing}.

\begin{figure*}[!tt]
	\centering
	\hspace{2em}\includegraphics[width=0.9\textwidth]{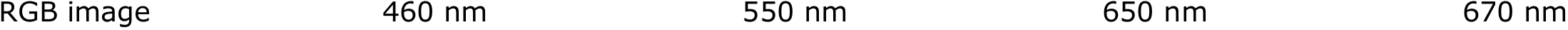}
	\\
	\begin{subfigure}[c]{0.015\textwidth}
		\centering
		\includegraphics[width=\textwidth]{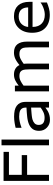}
	\end{subfigure}
	\begin{subfigure}[c]{0.97\textwidth}
		\centering
		\begin{tabular}{c}
			\animategraphics[width=\textwidth,loop]{5}{figures/candle/stack/im}{1}{50}
		\end{tabular}
	\end{subfigure}
	\\
	\begin{subfigure}[c]{0.015\textwidth}
		\centering
		\includegraphics[width=\textwidth]{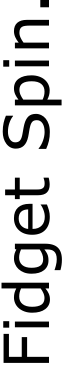}
	\end{subfigure}
	\begin{subfigure}[c]{0.97\textwidth}
		\centering
		\begin{tabular}{c}
			\animategraphics[width=\textwidth,loop]{5}{figures/fidget/stack/im}{1}{50}
		\end{tabular}
	\end{subfigure}
	\caption{\textbf{Video rate reconstruction.} We imagined challenging scenes such as a live flame, as well as a fidget spinner, both consisting of complex texture, spectrum, and temporal dynamics. Our system was able to accurately recover the HSIs at each time-frame.}
	\label{fig:video_master}
\end{figure*}

Figure \ref{fig:teaser} showed an example of two Matroshka dolls rotating on a turnstile. We showed the video sequence of three spectral bands over $100$ time frames (use Adobe Reader to view the animations).
%
%
The blue, and red colors on the dolls are distinctly visible in the spectral bands, along with accurate texture reconstruction.
Figure \ref{fig:video_master} shows two more video reconstructions. Both scenes show highly dynamic motions that are challenging to capture; our reconstructions indicate successful reconstructions of the hyperspectral video.
%
%
%

\begin{table}[!tt]
	\centering

	\caption{\small\textbf{Timing per frame.} We implemented a highly optimized pipeline for  video hyperspectral imaging. Shown below are the timing for the key steps in our acquisition pipeline. The numbers provided here were observed on a Dell Alienware computer with $9^{th}$ Gen Intel Core i7, 32GB RAM, and NVIDIA GeForce GTX 1660 Ti.}
	\label{tab:timing}
		\includegraphics[width=\columnwidth]{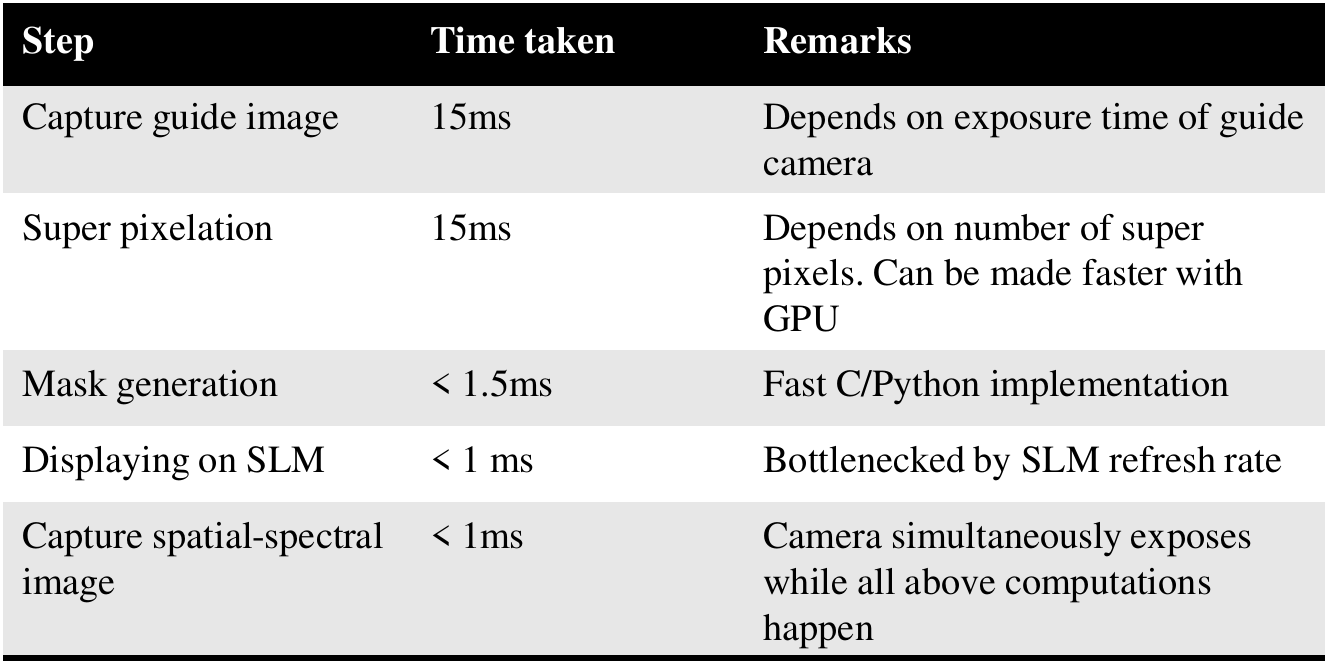}
\end{table}

\begin{figure*}[!tt]
	\centering
	\includegraphics[width=\textwidth]{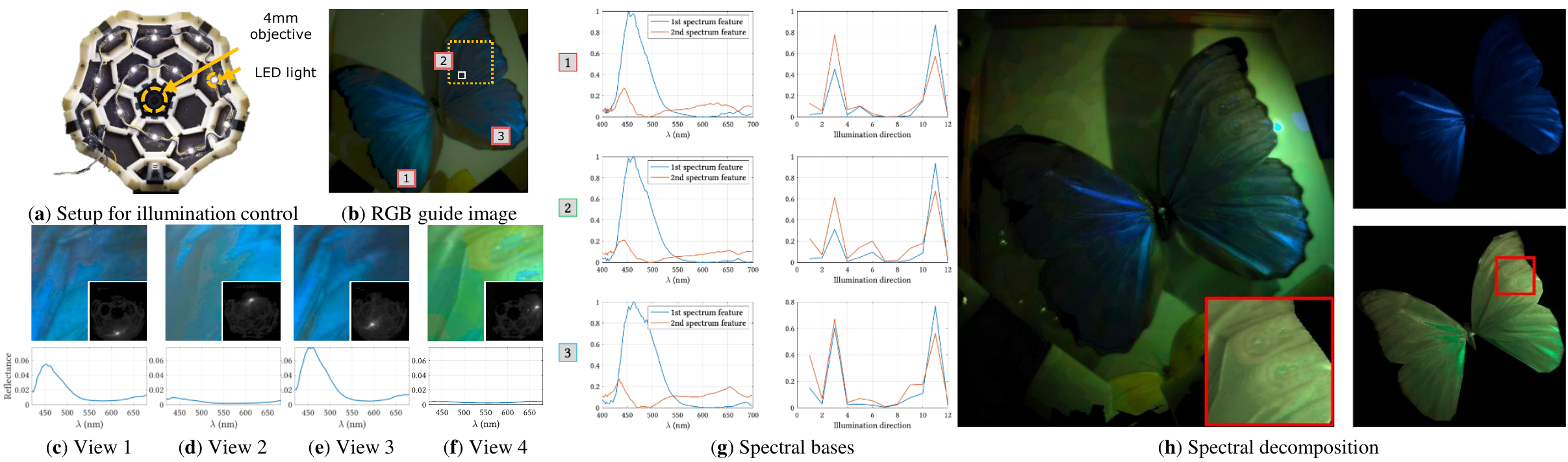}
	\caption{\textbf{Spectral BRDF.} We combined our setup with a light dome to estimate spectral BRDF of a butterfly, illuminated in different directions by a white LED light source. We then captured  HSIs for each illumination condition (see inset image for direction of illumination) and estimated spectrum on a small patch. We observed that the spectrum is dependent on illumination condition, as evident from deep blue color in (e) view 4 and brown shade in (f) view. (g) shows a plot of rank-2 decomposition of spectra at the three marked points, with the blue plot representing the iridescent blue spectrum, and the red plot representing the background, broadband spectrum. A spectral and angular decomposition of the object enabled us to separate the scene into iridescent and non-iridescent components, all achieved with only one image captured per angle.}
	\label{fig:spectral_brdf}
\end{figure*}

\bpara{Spectral reflectance.}
%
%
SASSI provides the ability to get HSIs at the specifications of a pushbroom device, but without the long scanning times.
This is very useful when we want to measure higher dimensional slices  of the plenoptic function, for example, reflectance hyperspectral fields to study iridescene and structural coloration, which are  spectral phenomena.

%
The setup used for this purpose is shown in Fig. \ref{fig:spectral_brdf} (a), which consists of the SASSI camera  and a dome of twelve white LED sources well spread over the hemisphere.
We imaged a butterfly, which exhibits rich structural colorations as the angle of illumination changes.
%
%
The spatial images as well as spectral reflectances are shown in Fig. \ref{fig:spectral_brdf} (c) through (f). 
Certain illumination conditions, such as view 2, 4 showed strong blue peaks, whereas view 5 produced a brown colored image, which reveals the eye-like structure under the wing.

Using this data, we decomposed the measured spectrum into a basis that separated the blue iridescent reflectance from the brown underlying reflectance using non-negative matrix factorization.
%
%
This provided us with a set of spectral feature vectors and corresponding vector weights combining these features at every point.
%
We first applied this only to the spectral dimension, with a 2-dimensional decomposition.
The results are shown in Fig. \ref{fig:spectral_brdf} (g).
We can observe that the dominant spectral features consist of a high intensity blue specrum, and a low intensity brown spectrum.
%
%
%
%
%
These angles match the view points in Fig. \ref{fig:spectral_brdf} that exhibit deep blue color and are consistent across different parts of the butterfly that show iridescence at the same angles.

We then decomposed the HSIs along spatial and illumination directions.
By separating these features into iridescent and non-iridescent components, we were able to decompose an image into two separate images corresponding to these components.
The results of this decomposition are shown in Fig. \ref{fig:spectral_brdf} (h).
Using only a few measurements, we are were to separate the blue iridescent reflectance from the brown underlying reflectance, which reveals the patterns on the butterfly's wing.

\section{Conclusion} \label{section:conclusion}
%
%
%
%
%
Our paper shows that adapting the sampling patterns to the specific instance of a scene can significantly reduce measurement time, without sacrificing spatial, or spectral resolutions.
By making hyperspectral imaging faster and robust, SASSI opens up novel applications in  material identification, biological imaging, and computer vision tasks.
It also opens up the possibility for measuring higher dimensions of light involving spectrum, angle and polarimetry that has hitherto been hard to sense.
%
Hence, we believe the ideas presented in this paper will push the boundaries of plenoptic imaging and its associated applications.

\section{Acknowledgment}\label{sectiion:ack}
The authors acknowledge support via the National Geospatial Intelligence Agency's Academic Research Program (Award No. HM0476-17-1-2000), the NSF SaTC award 1801382, and the NSF Expeditions award 1730147.

\begin{appendices}
\section{Overview}\label{section:sup_introduction}
This article supplements the main paper and provides the following details:
\begin{itemize}
	\item \textit{Reconstruction details} gives an in-depth analysis of  computational complexity for rank-1 reconstruction, and learning details, and ablation studies for the learned guided filtering approach.
	\item \textit{Calibration} provides detailed steps to calibrate the SASSI optical system.
	\item \textit{Experiments with lab prototype} provides extra results with SASSI lab prototype with emphasis on variety of experiments, along with an example of failure mode.
\end{itemize}

\subsection{Reconstruction Details}\label{section:sup_nn}
We provide details about computational complexity of rank-1 estimate, as well as the architecture for learnable guided filtering.
\subsubsection{Computational complexity for rank-1 estimate.} We now provide an estimate for complexity of reconstruction to illustrate its efficacy.
Consider an HSI of size $N_x \times N_y \times N_\lambda$. Let $Q$ be the number of pixels at which spectral samples are measured.
Then reconstruction using the rank-1 approach requires averaging the sampled spatial locations in each super-pixel and appropriately scaling and assigning to non-sampled locations.
This can be efficiently performed by creating a dictionary where the keys are super-pixel numbers, and the entries consist of all pixel locations belonging to that super-pixel.
Generating such a dictionary requires $N_x N_y$ computations, while using the dictionary to average and assign spectrum on the $l^\textrm{th}$ super-pixel requires $N_l$ computations where $N_l$ is the number of pixels within the super-pixel.
In total, the complexity is $N_x  N_y + \sum_{l=1}^Q N_l = \mathcal{O}(N_x  N_y)$, implying that the reconstruction requires as many computations as the number of pixels in an image.
Compared to solving a complex optimization problem, our approach is evidently light weight both on computation and memory requirements.

\subsubsection{Guided filtering}
We now explain how guided filtering can be efficiently implemented using a series of convolutions.
\bpara{Efficient guided filtering approach.}
Each spectral band can be represented as, $I_\lambda(x, y) = M(x, y)\odot H(x, y, \lambda)$, where $M(x, y)$ is the spatial mask.
We can write guided filtering equation as,
\begin{align}
	I^k_\lambda \approx \alpha^k_\lambda I^k_\text{guide} +\beta^k_\lambda.
\end{align}
Since observations are made only where $M(x, y)$ is non-zero, we modify the optimization problem as,
\begin{align}
	\min_{\alpha^k_\lambda, \beta^k_\lambda} \left\|M\odot\left(I^k_\lambda - \alpha^k_\lambda I^k_\text{guide} -\beta^k_\lambda\right)\right\|^2,
\end{align}
which is a simple weighted least squares optimization problem.

An efficient, closed form solution \cite{wu2018fast} for the above equation  is the following,
\begin{align}
	\alpha(x, y) = \frac{\mu_{xy} - \mu_x \mu_y}{\mu_{xx} - \mu^2_x + \epsilon},\ \ \ 
	\beta(x, y) = \mu_y - \alpha(x, y)\mu_x \\
	I_\lambda(x, y) = \alpha(x, y)I_\text{guide} + \beta(x, y)\\
	\mu_x = \text{boxfilter}(I_\text{guide}(x, y) M(x, y), N)/C_M\\
	\mu_y = \text{boxfilter}(I_\lambda(x, y) M(x, y), N)/C_M\\
	\mu_{xx} = \text{boxfilter}(I^2_\text{guide}(x, y) M(x, y), N)/C_M\\
	\mu_{xy} = \text{boxfilter}(I_\text{guide}(x, y)\cdot I_\lambda(x, y) M(x, y), N)/C_M\\
	C_M = \text{boxfilter}(M(x, y), N),
\end{align}
where $\text{boxfilter}(\cdot, N)$ is a 2D uniform filter of size $N$, and $\epsilon$ is a stabilization constant.
%

\begin{figure}[!tt]
	\centering
	\includegraphics[width=\columnwidth]{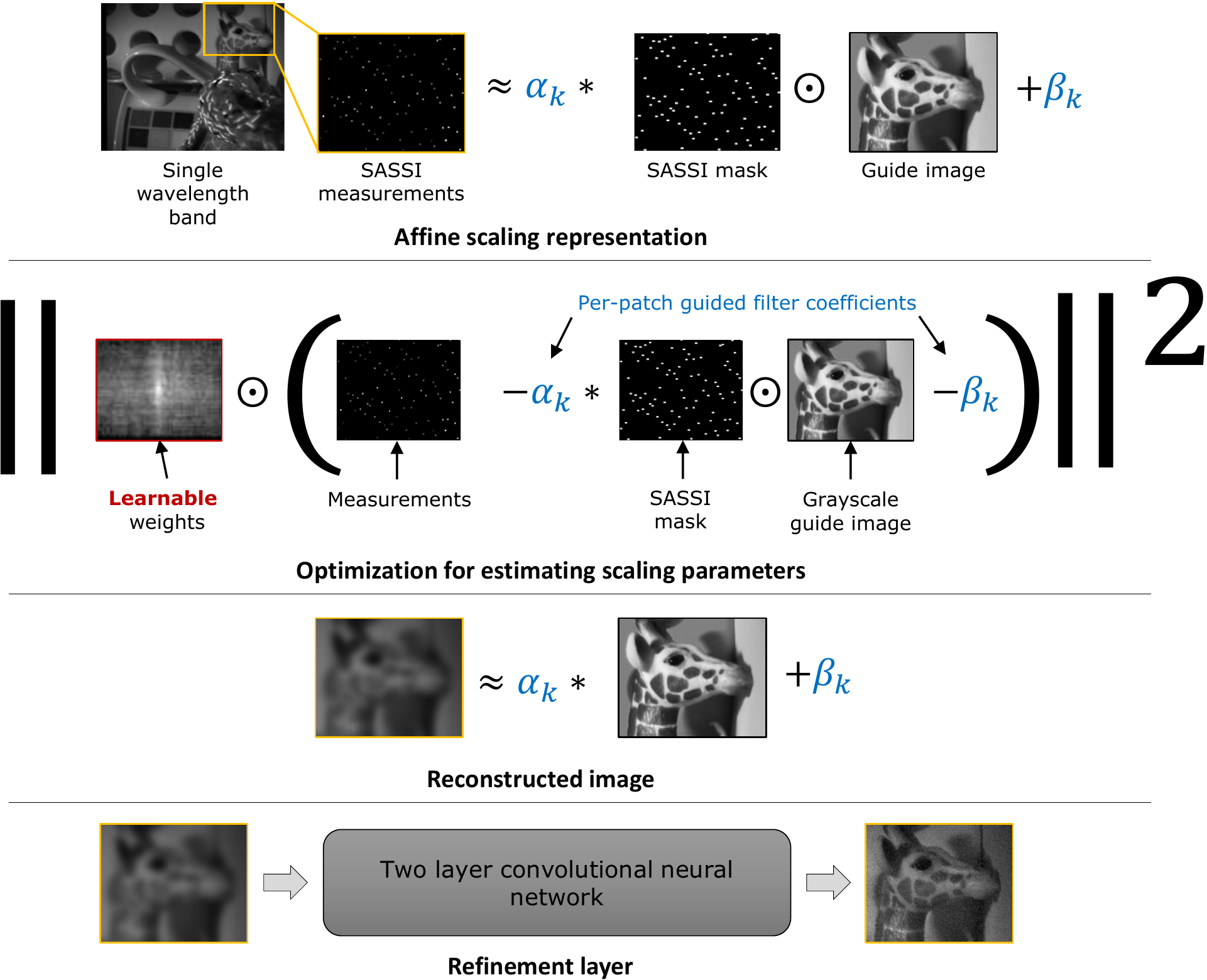}
	\includegraphics[width=\columnwidth]{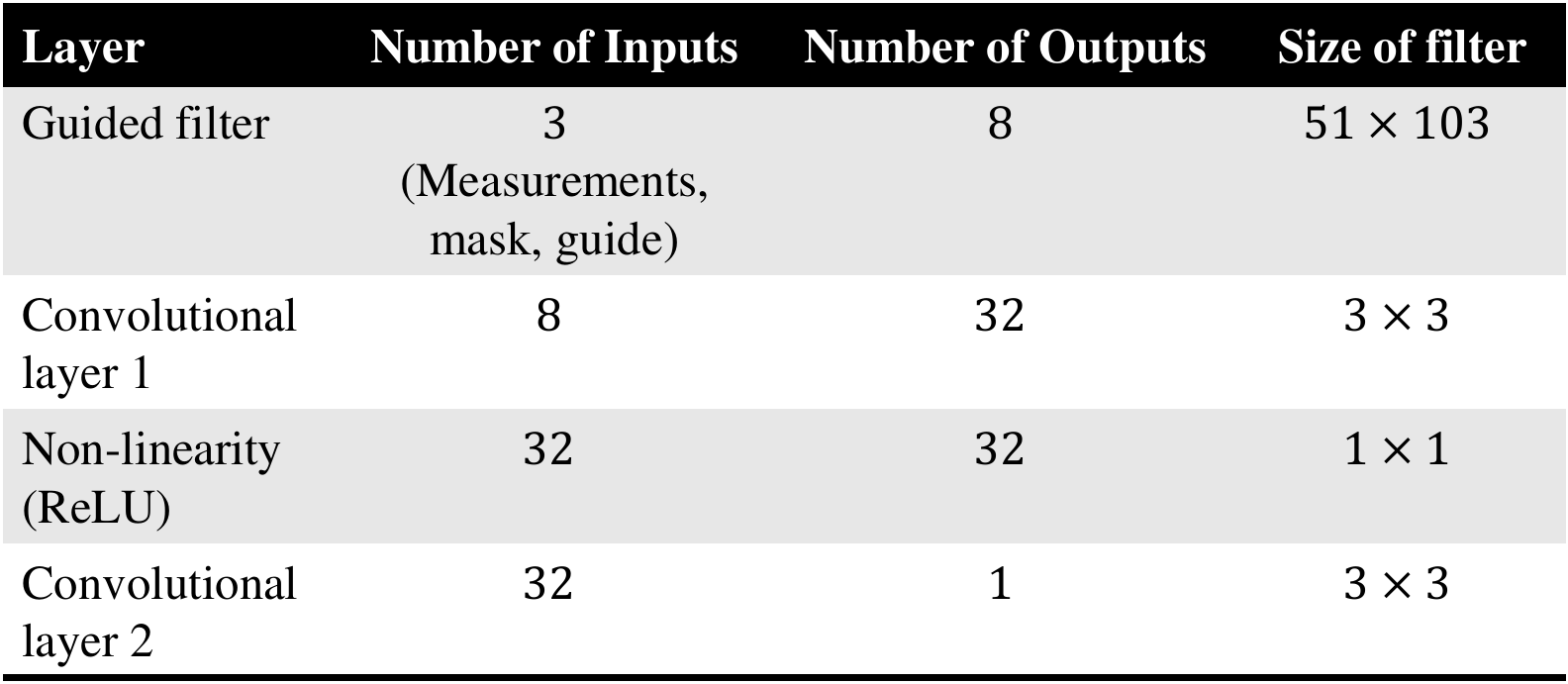}
	\caption{\textbf{Visualization of guided filtering layer.} We rely on a simple modification to traditional guided filtering. We represent the measurements as an affine scaling of the grayscale guide image. Then we solve a \textit{weighted} least squares problem to estimate the coefficients, which is then used to reconstruct image at each wavelength band. This is followed by a two-layer refinement stage to get the final output. The learnable weights are trained along the two-layer neural network with a composite loss consisting of SSIM and MSE loss functions. The table above shows dimensions of each layer.}
	\label{fig:nn_model}
\end{figure}

Instead of a fixed, uniform filter, we rely on \emph{learnable} stack of 2D filters to reconstruct each spectral band.
Such a per-band approach, instead of a full 3D convolution approach ensures that the reconstruction does not depend on the smoothness of the spectrum itself, and hence can be applied to varying set of scene and illumination conditions.
Figure \ref{fig:nn_model} provides a visualization of the guided filtering approach.

\bpara{Choice of network parameters.}
The hyperparameters of the reconstruction network consists of number of guided filters, size of each filter, number of refinement layer, and size of each refinement layer.
The details of the parameters are provided in Fig. \ref{fig:nn_model}.
We note that several modifications of this architecture are possible, and potentially capable of giving higher accuracy, however, we found this simple three layer architecture to be effective.
We trained two separate neural networks in our paper, one for simulations and one for real results, but followed similar design principles.
We explain the choices for simulations here; the extension to real experiments will be briefly mentioned.
We briefly discuss our motivation for filter sizes and number of filters.

\begin{figure}[!tt]
	\centering
	\includegraphics[width=\columnwidth]{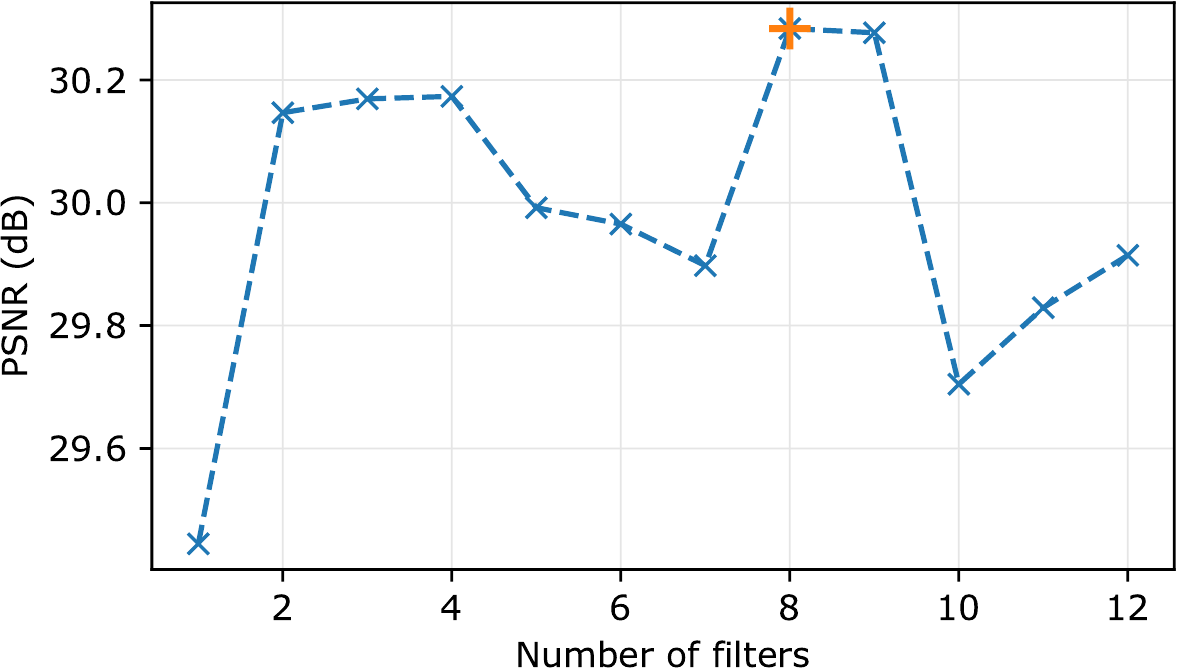}
	\caption{\textbf{Accuracy with varying number of filters.} We varied the number of filters from $1$ to $12$ with a fixed filter size of $31\times63$ and found that $8$ filters gave the optimal results. We hence used $8$ filters in all our experiments.}
	\label{fig:nfilters}
\end{figure}

\bpara{Number of filters.}
It is possible, and in practice advantageous to have multiple guided filters in the first layer.
Ideally, fewer filters prevents overfitting and has smaller computational complexity.
In contrast, a larger number of filters achieves better reconstruction accuracy.
To choose the optimal number of filters, we trained our neural network with varying number of filters, and a fixed filter size of $31\times63$.
Figure \ref{fig:nfilters} shows the plot of accuracy as a function of number of filters for the KAIST dataset.
We observed the peak accuracy to be at $8$ filters, which was the number of filters we used in all our experiments.

\begin{figure}[!tt]
	\centering
	\includegraphics[width=\columnwidth]{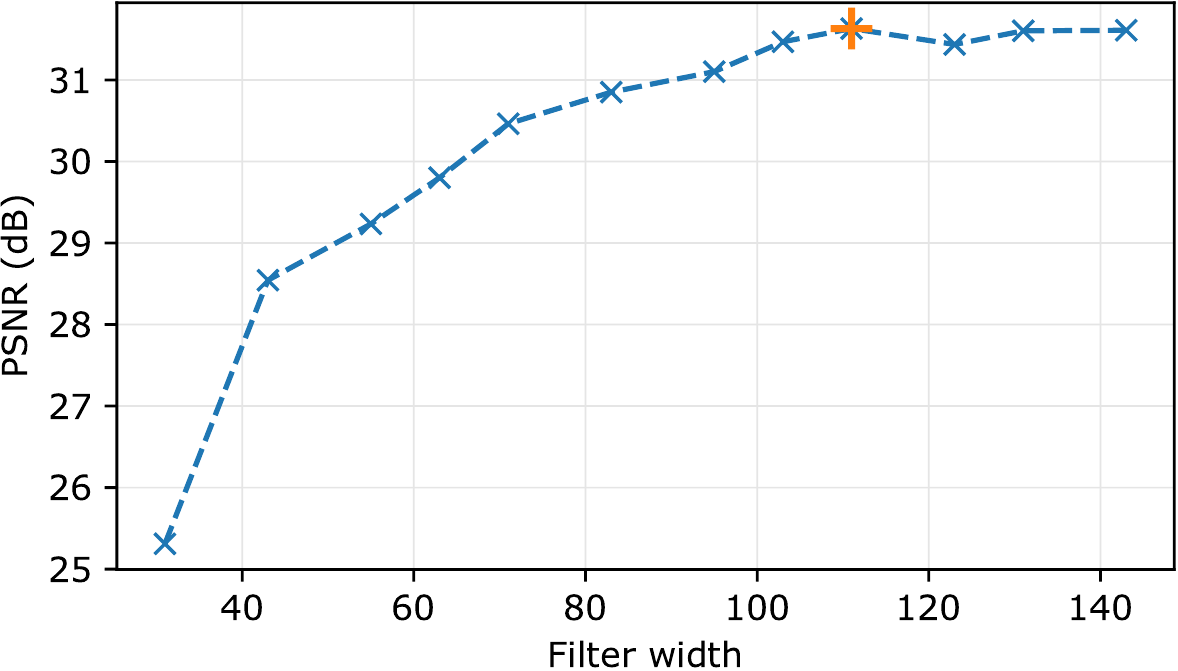}
	\caption{\textbf{Accuracy with varying filter size.} We varied the filter width from $31$ to $150$ with the height fixed to half the width. The number of filters were fixed to $8$. We observed that the accuracy peaked between $101 - 111$. We chose a slightly smaller filter size, as we found it provided a small gain in speed.}
	\label{fig:nbands}
\end{figure}

\bpara{Size of filters.}
Our simulation datasets consisted of images with $31$ spectral bands.
To ensure that every patch gets at least one measurement, we need to ensure that the \textit{width} of the guided filter is at least $31$ dimensional, with the height being variable.
In practice, we found that filters sizes slightly bigger than $31\times31$ gave better results, particularly under noisy measurements.
Figurer \ref{fig:nbands} shows accuracy as a function of filter size, with filter height being half the width.
We found a filter width between $101 - 111$ gave best results over simulations and real experiments.
We chose $101$ filter size as it provided gains in speed.

\bpara{Ablation studies}
To evaluate the importance of each individual component, we performed ablative studies by evaluating different variations of our mode, the details are listed in Tab. \ref{tab:ablation}.
We evaluated a total of five models:
\begin{itemize}
	\item \textit{Naive guided filtering with box filter} consisted of a single guided filtering layer with a uniform box filter.
	\item \textit{Naive guided filtering with gaussian filter} consisted of a single guided filtering layer with a gaussian shaped filter.
	\item \textit{Learned guided filter only} consisted of a single guided filtering layer that was learned on the data.
	\item \textit{Refinement only} consisted of a fixed guided filtering layer (Gaussian shaped) along with the two-layer neural network.
	\item \textit{Learned guided filter and refinement} consisted of $8$ guided filters along with refinement, all learned together..
\end{itemize}
In each case, we learned the model over $100$ epochs with $10,000, 256\times256$ 2D spectral images, at a learning rate of $10^{-4}$.
We tested our model on the ICVL dataset~\cite{arad_and_ben_shahar_2016_ECCV}, and evaluated PSNR, and SSIM.
We observe that the model with all parameters learned together outperformed in both PSNR and SSIM, thus validating the choice of our model.
\begin{table}[!tt]
	\caption{\small\textbf{Ablative studies for choice of our model.} Training the model parameters including the guided filter layer gave the best performance, which validated our neural network choice.}
	\label{tab:ablation}
	\includegraphics[width=\columnwidth]{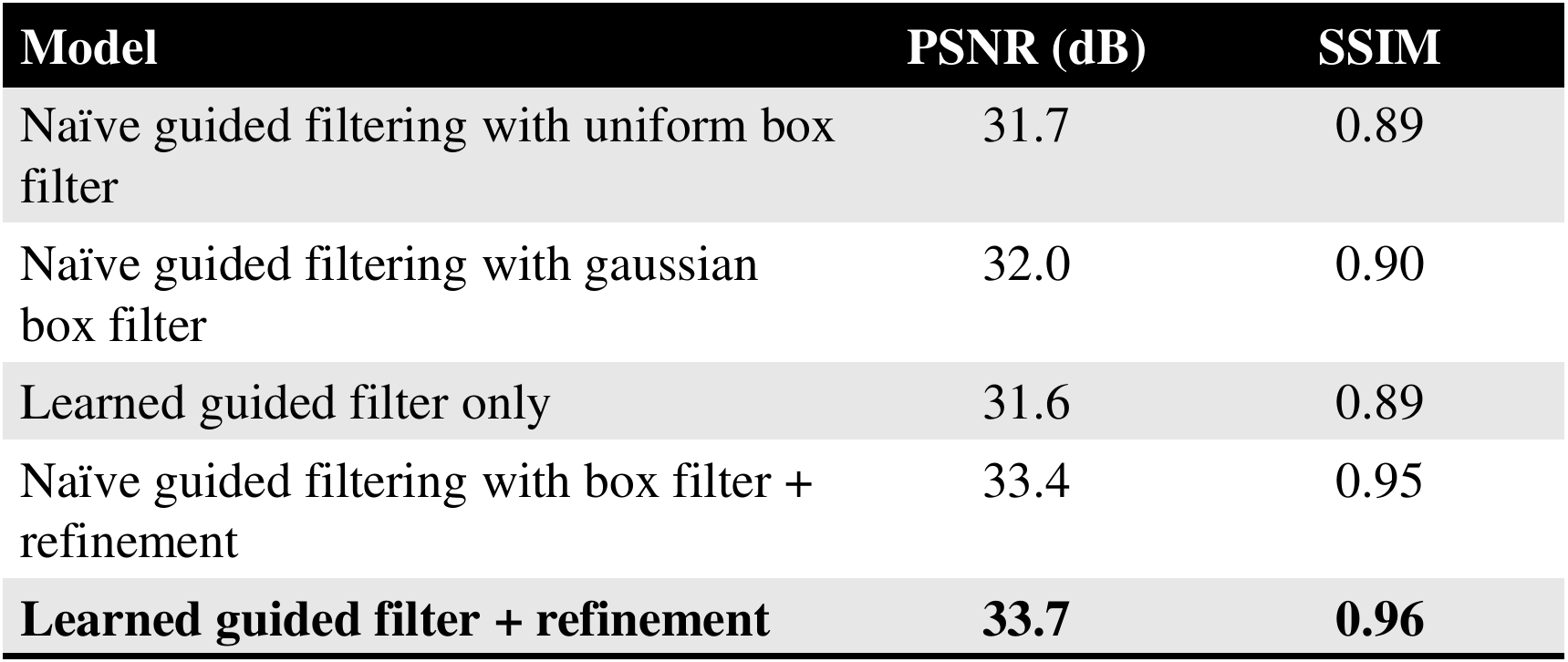}
\end{table}

\bpara{Training details.}
We picked the KAIST dataset~\cite{deepcassi2017} as it contained several scenes with interesting spectral and spatial textures.
We trained our neural network with a composite loss function consisting of Structural Similarity (SSIM), and mean squared error (MSE), with weights, $\eta_\text{SSIM} = \eta_\text{MSE} = 1$.
We found that initializing the guided filtering layer with Gaussian-shaped filters lead to faster convergence.
We trained our models on PyTorch with a learning rate of $\eta = 10^{-4}$ and ADAM optimizer with $\beta_1 = 0.9, \beta_2 = 0.999$ for a total of $1,000$ epochs.
We used a total of $100,000, 256\times256$ 2D spectral images of which $50,000$ was used for training and $50,000$ for testing and validation.

\begin{figure}[!tt]
	\centering
	\includegraphics[width=\columnwidth]{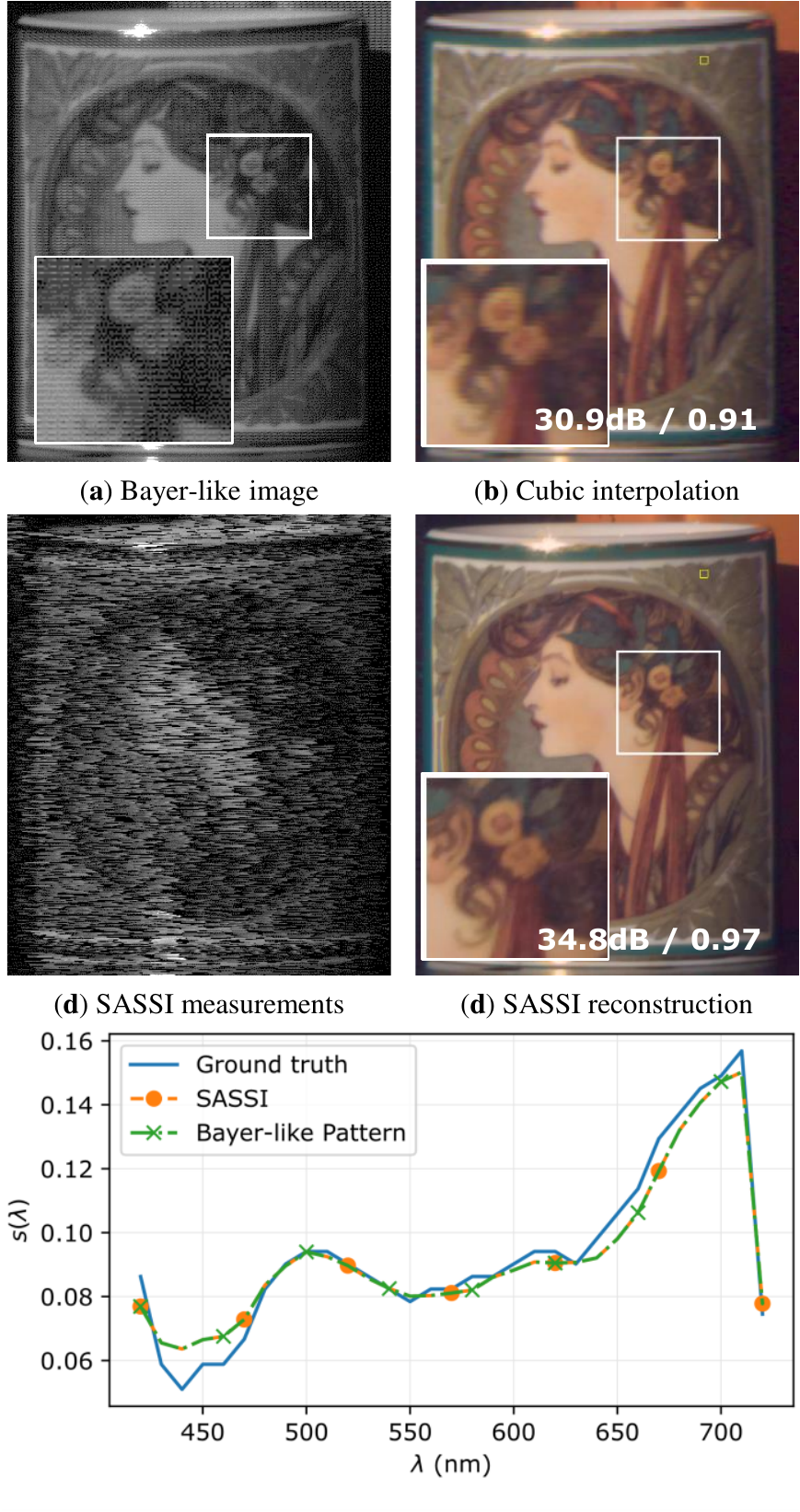}
	\caption{\textbf{Comparison against Bayer-like sampling.} Bayer-like filter patterns provide snapshot hyperspectral imaging ability. While their spectral fidelity is high, spatial resolution is inferior compared to SASSI.}
	\label{fig:bayer}
\end{figure}

\bpara{Bayer-like snapshot hyperspectral sensor.}
A promising snapshot HSI sensor design is a Bayer-like pattern of spectral filters on a grayscale sensor, similar to the commercially available products by IMEC\footnote{\url{https://drupal.imec-int.com/sites/default/files/inline-files/Mosaic\%20SWIR\%203x3\%20snapshot\%20SWIR\%20range\%20hyperspectral\%20imaging\%20camera_0.pdf}}.
To compare SASSI and Bayer-like snapshot cameras, we generated a grayscale imaging with periodic tiling of $4\times8$ filters for 32 spectral bands, as shown in Fig. \ref{fig:bayer} (a).
We then reconstructed the full image from the grayscale image using a simple, cubic interpolation, shown in \ref{fig:bayer} (b).
We then simulated a SASSI measurement shown in Fig. \ref{fig:bayer} (c), and then reconstructed using a rank-1 approach shown in \ref{fig:bayer} (d).
While the spectral accuracy is high, the spatial resolution of Bayer-like sensor is inferior to SASSI -- this is evident from reconstruction SNR and SSIM.

\subsection{Calibration} \label{section:sup_calibration}
\begin{figure}[!tt]
	\centering
	\includegraphics[width=\columnwidth]{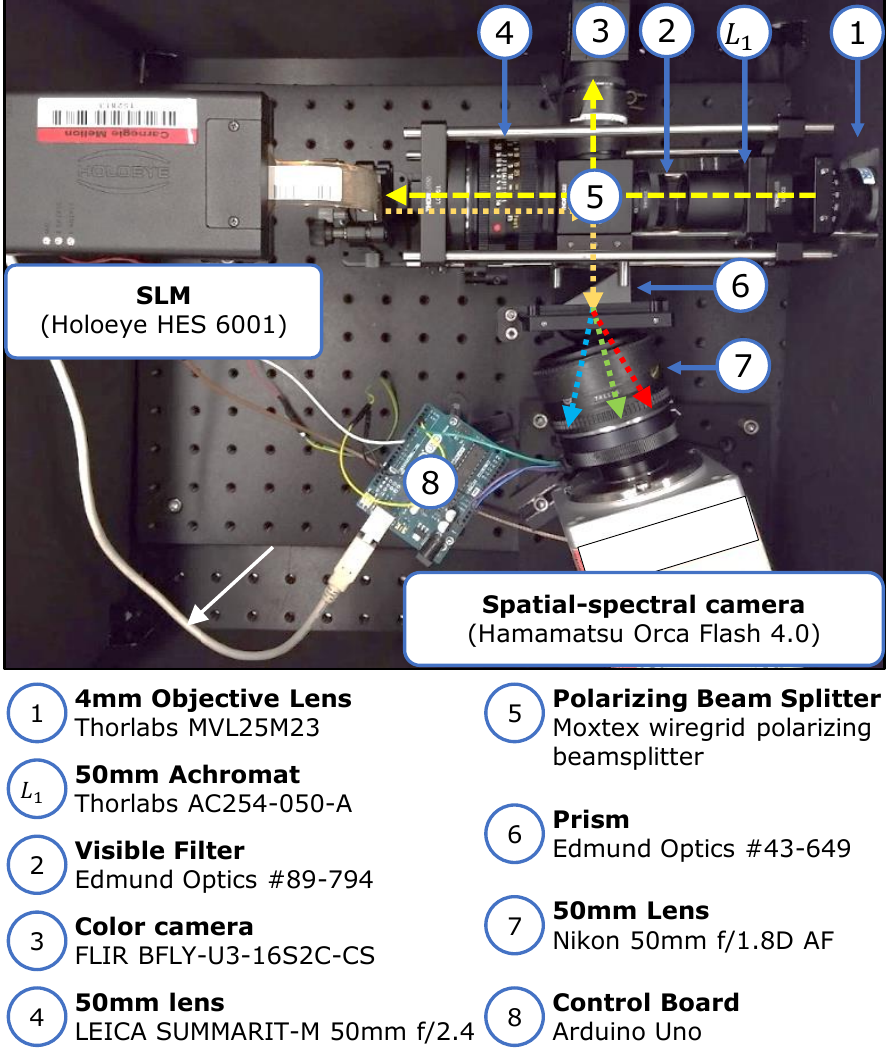}
	\caption{\textbf{List of components.}}
	\label{fig:components}
\end{figure}

We first provide list of components for our setup and the steps taken to calibrate the system.
\bpara{Components list.}
Figure \ref{fig:components} lists out all the major components used for building the SASSI prototype. 
The RGB and grayscale cameras were triggered by a master pulse from the SLM.
The Arduino board was used to filter the pulse to enable capture only during the exposure duration.
We used an off-the-shelf objective lens in front of the SLM (Leica 50mm) to get sharper image with minimal spectrally-dependent blur.

\subsubsection{Calibration}
\begin{figure}
	\centering
	\begin{subfigure}[c]{\columnwidth}
		\centering
		\includegraphics[width=\columnwidth]{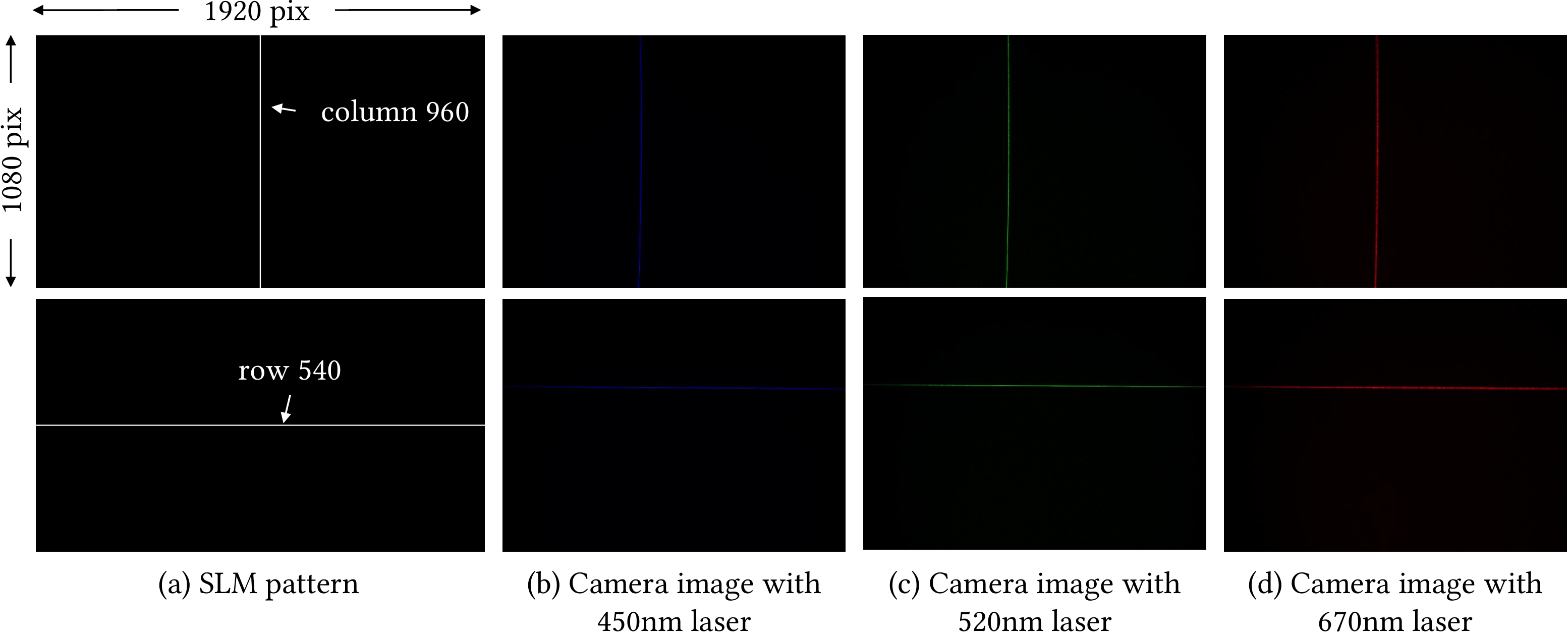}
		\caption{Calibration illustration}
	\end{subfigure}
	\begin{subfigure}[c]{0.48\columnwidth}
		\centering
		\includegraphics[width=\columnwidth]{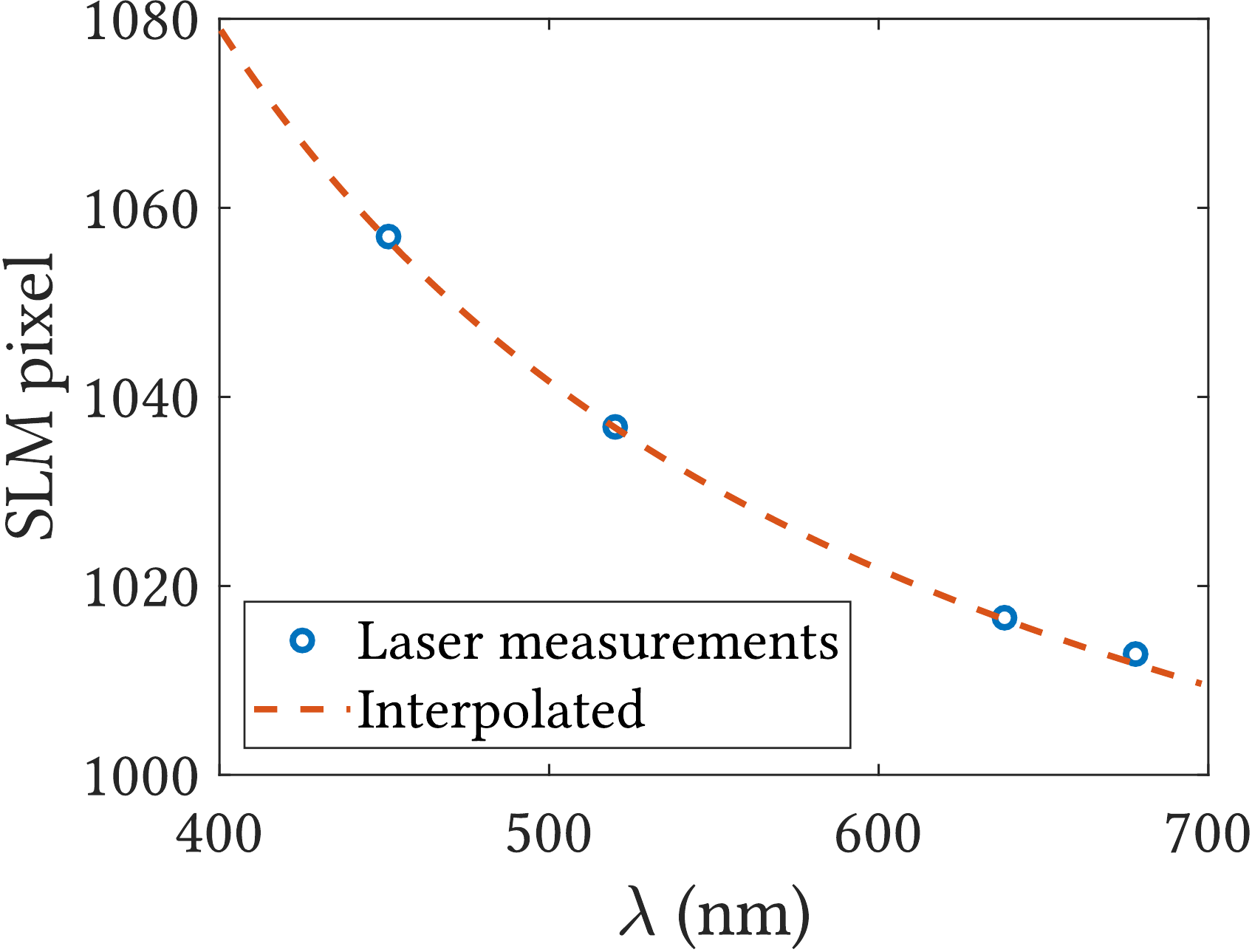}
		\caption{Interpolation of column correspondences}
	\end{subfigure}
	\begin{subfigure}[c]{0.48\columnwidth}
		\centering
		\includegraphics[width=\columnwidth]{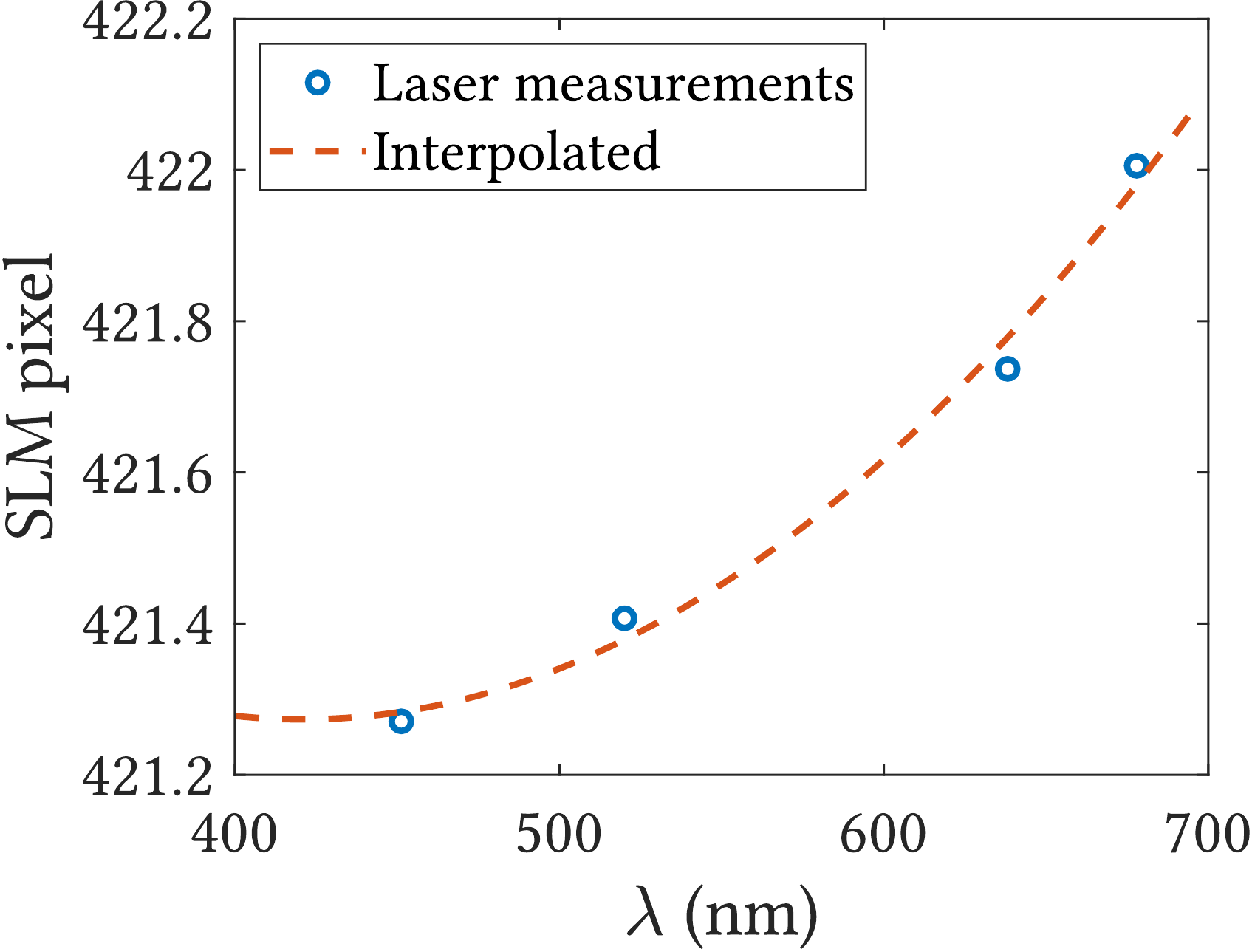}
		\caption{Interpolation of row correspondences}
	\end{subfigure}
	\caption{\textbf{Calibrating SLM and spatial-spectral camera.} We performed a pixel-to-pixel mapping from SLM to spatial-spectral camera by turning on one column/row at a time for five different lasers (450, 520, 532, 635 and 670 nm) and finding the corresponding pixels in the camera. We then used the correspondences obtained with lasers to interpolate for other wavelengths. For (b) column correspondences, we assumed that the SLM coordinates were a second degree polynomial function of the prism's refractive index, while for (c) row correspondences, we used a second degree polynomial of wavelengths.}
	\label{fig:slm_calib}
\end{figure}

Our system requires calibrating images to a common coordinate 2D system.
This involved calibrating the SLM and spatial-spectral camera, and the SLM and the guide camera.
To calibrate the SLM and spatial-spectral camera, we performed a pixel-to-pixel mapping between the two devices.
This was achieved by uniformly illuminating a white object with a narrow band laser, and then switching on one column/row of the SLM at a time and capturing images.
Illustration of the SLM to spatial-spectral camera calibration is shown in Fig. \ref{fig:slm_calib} (a).
We used a total of 5 lasers at $450, 520, 532, 635$ and $670$ nm to obtain five column/row correspondences.
We then interpolated to other wavelengths by a smooth interpolation method. For column correspondences, we assumed that the indices were a second degree polynomial function of the refractive index of the prism for each wavelength. We found this gave the best fit for the spread of wavelengths.
For interpolating row correspondences, we assumed that the indices were a second degree polynomial of the wavelengths.
The measured SLM camera correspondences, as well as interpolated values are shown in Fig. \ref{fig:slm_calib} (b) and (c).
The output of this calibration step was $68$ pixel-to-pixel correspondences for SLM and spatial spectral camera.

\begin{figure}[!tt]
	\centering
	\includegraphics[width=\columnwidth]{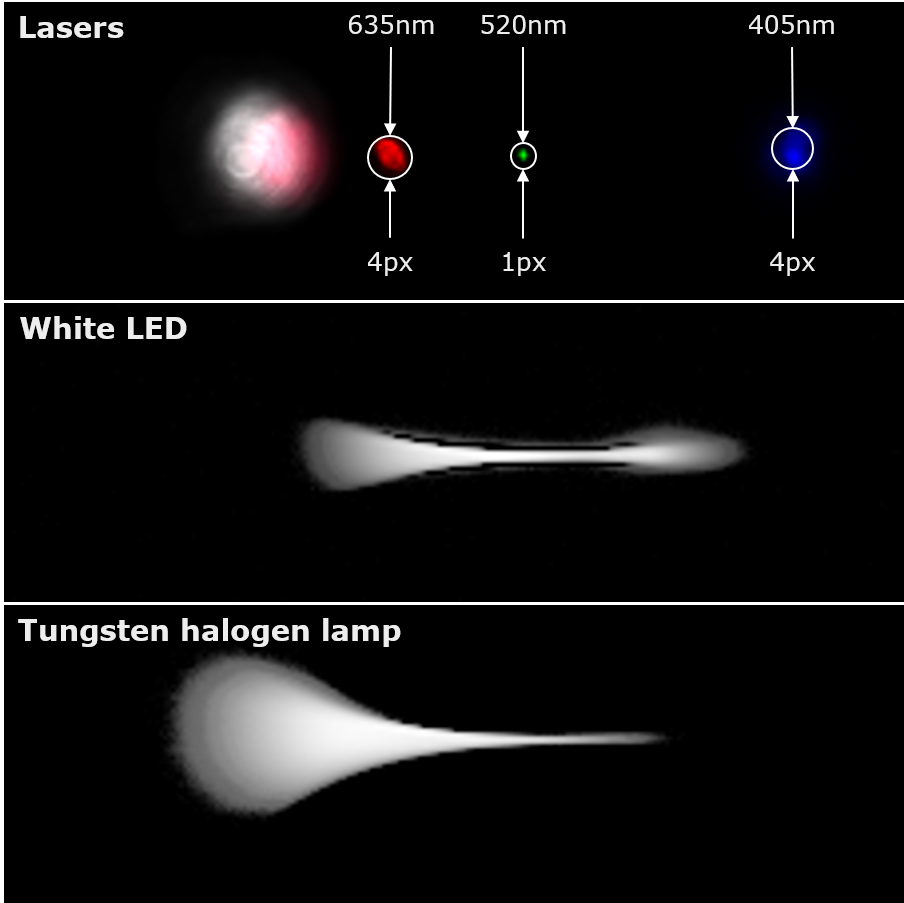}
	\caption{\textbf{Wavelength-dependent blur.} We observed that the optical system produced wavelength-dependent blur. We compensate for the optical blur by fitting a second-order polynomial to the blur kernel size, with a blur of 4 pixels at 405, and 635nm, and 1 pixel at 520nm.}
	\label{fig:wavelength_blur}
\end{figure}

\bpara{Wavelength-dependent blur.} While we utilized the best possible lenses for our optical setup, there was invariably a wavelength-dependent blur in the measured spectral profiles.
Prior work on CASSI-based optical systems largely ignore this aberration, which leads to inaccuracies in reconstructions. 
Since our measurement relies on capturing non-overlapping spectra, we can compensate by estimating the blur kernel size for each wavelength.
To achieve this, we estimate the blur kernel at each wavelength, and then sum up the signal under the area of the blur kernel.
Figure \ref{fig:wavelength_blur} visualizes the blur as a function of wavelength for various light sources.
We observed that the spectral blur was 4 pixels for 405, and 635nm, and 1 pixel for 520nm. 
This is consistent with the fact that most achromatic doublets have a focal length shift that varies quadratically with wavelength.
Hence we fitted a quadratic function to the observed blurs at three different wavelengths.
Further, to avoid overlap of spectral streaks, we added a spacing of 5 pixels vertically.
This reduced the light throughput of the system, but ensures high fidelity reconstruction of spectra.

\begin{figure}
	\centering
	\includegraphics[width=\columnwidth]{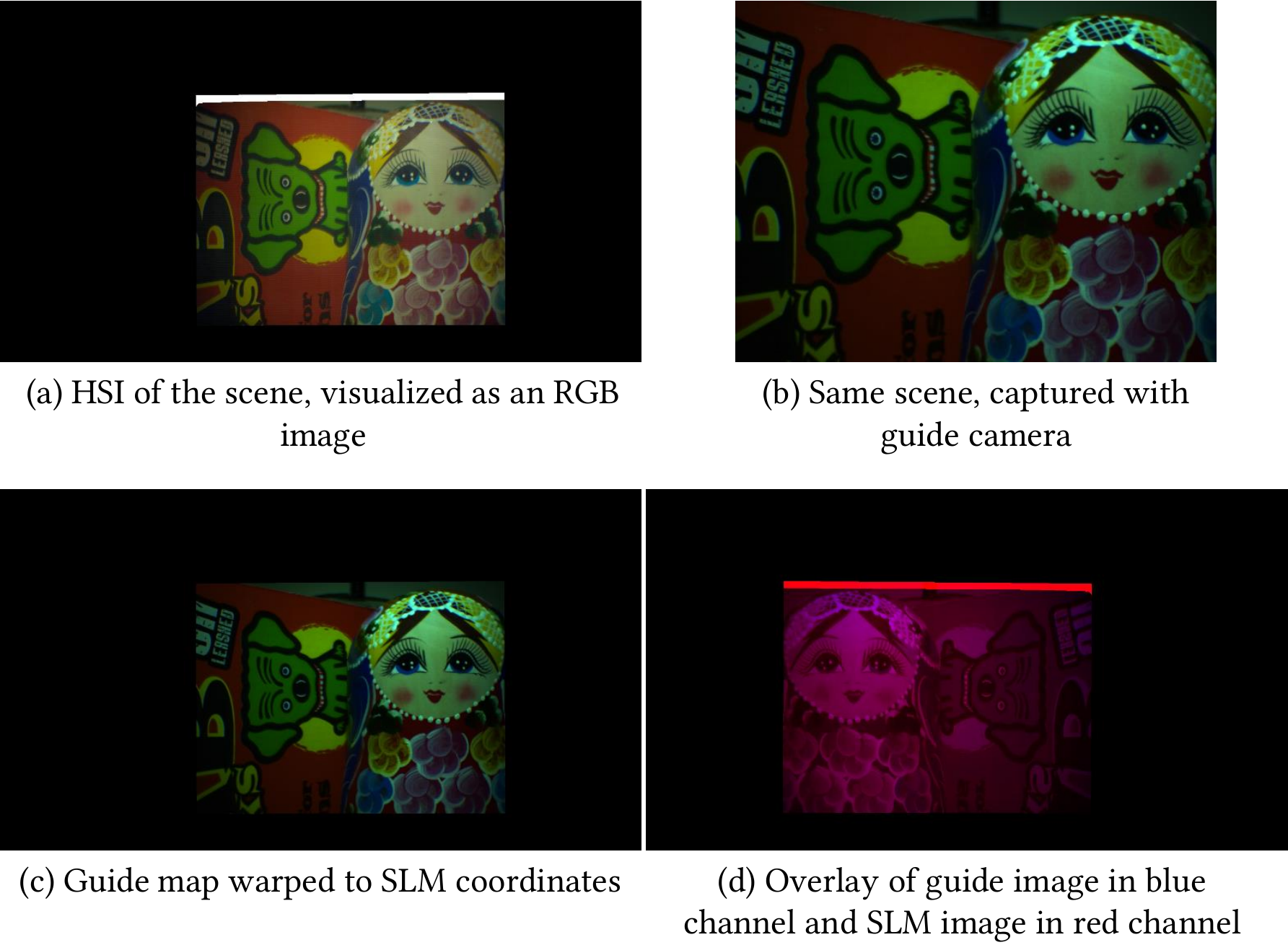}
	\caption{\textbf{Calibration of SLM and guide camera.} We captured a highly textured scene by scanning (a) the full HSI and mapping it to the SLM, and (b) capturing an image with the guide camera. We then assumed a projective transformation between the two images using SURF features \cite{bay2006surf}, which allowed us to accurately warp the guide image to (c) SLM coordinates. (d) shows an overlay of the two images, showing the accuracy of our registration.}
	\label{fig:guide_calib}
\end{figure}

\bpara{Guide camera calibration.}
To get correspondences between SLM and guide camera for calibration, we followed an indirect method which involved the spatial-spectral camera.
We first scanned the HSI of a scene (we describe the details of fully scanning later) with a lot of texture, and generated an RGB image as would be seen on the SLM.
Simultaneously, we captured an RGB image with the guide camera.
Since the SLM and camera are optically colocated, we modeled the transformation between the two RGB images as a simple projective transformation. using SURF features \cite{bay2006surf}. 
The output of this step was a $3\times3$ matrix that mapped guide image to the coordinates of the SLM.
Figure \ref{fig:guide_calib} shows an illustration of the calibration procedure.

\subsection{Experiments with Lab Prototype} \label{section:sup_real}
This section presents some more experiments with our lab prototype to demonstrate the versatility of SASSI.
We also present a failure case to show the limits of the system.

\begin{figure*}[!tt]
	\centering
	\includegraphics[width=\textwidth]{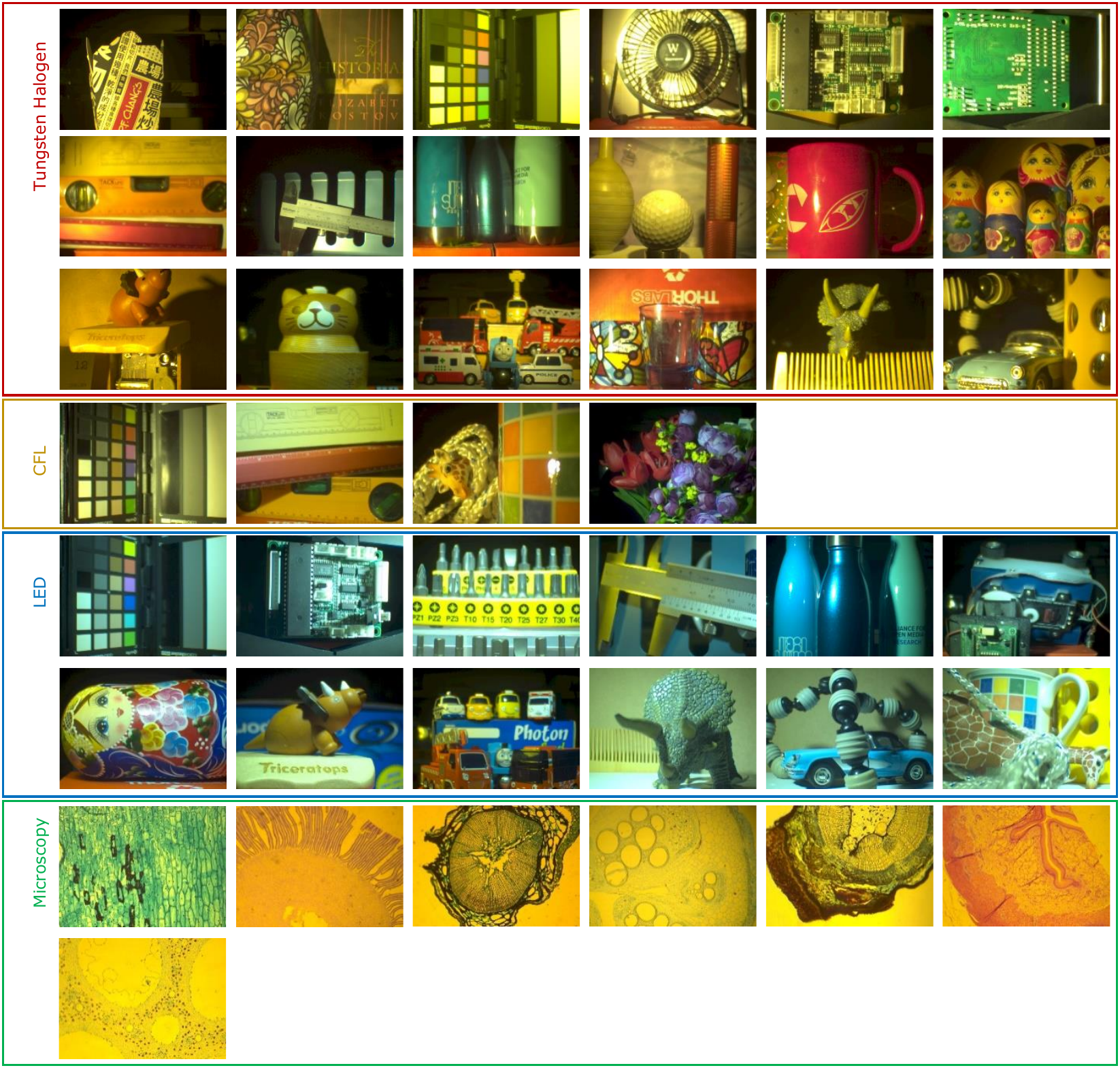}
	\caption{\textbf{Dataset visualization.} We collected datasets with varying illumination conditions and of a variety of objects. We included seven microscopic datasets. We will be making our dataset public to facilitate further research.}
	\label{fig:dataset}
\end{figure*}
\begin{figure*}[!tt]
	\centering
	\includegraphics[width=0.8\textwidth]{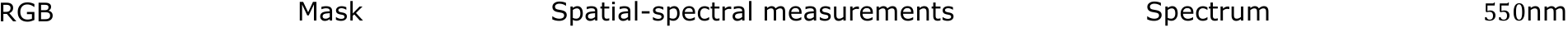}
	\begin{tabular}{c}
		\animategraphics[width=\textwidth,loop]{5}{figures/catspin/specstack/im}{1}{50}
	\end{tabular}
	\caption{\textbf{Video reconstruction visualization.} The video sequence can be viewed in Adobe Reader. Each frame shows RGB image, the mask, the captured CASSI image, spectra at the marked point and reconstruction at $550$nm. Notice how the spectrum magnitude reduces, and shifts towards red, as the point moves from white to brown.}
	\label{fig:catspin}
\end{figure*}
\begin{figure*}[!tt]
	\centering
	\includegraphics[width=\textwidth]{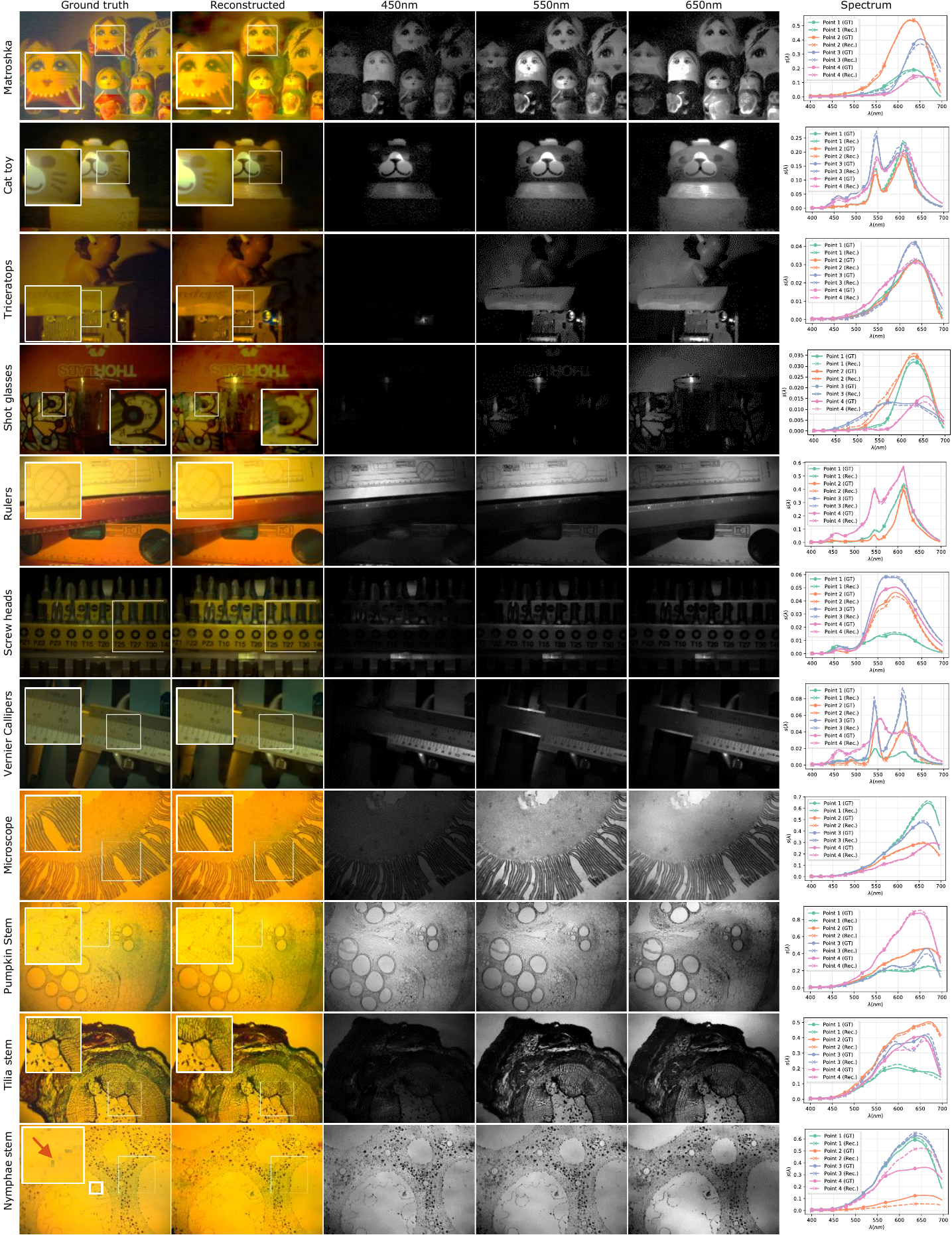}
	\caption{\textbf{Real experiments.} Reconstruction was performed with guided filtering approach. None of the scenes overlapped with the training set.}
	\label{fig:sup_real_master}
\end{figure*}

\bpara{Dataset.}
We collected several hyperspectral images with several spectrally and spatially interesting objects, with varying light conditions.
Figure \ref{fig:dataset} shows a picture of all the images we captured.
Figure \ref{fig:catspin} shows an example of a cat toy spinning, with spectrum plotted at a fixed point as a function of time, and  Fig. \ref{fig:sup_real_master} visualizes some more real results with emphasis on complex illuminations such as compact fluorescent lamp (Cat toy, and Rulers) and mixed illumination conditions (Vernier Calipers).
We plan to release the images, and the video sequences for further research.
\bpara{Macbeth color chart.}
To measure radiometric accuracy of our optical system, we imaged the Macbeth color chart that consists of 24 swatches of colors.
We illuminated the scene with a white LED. We estimated reflectances by dividing the spectral profiles at each pixel by spectrum of LED.
We observe that our method estimates the spectra very accurately, establishing the efficacy of our optical setup.

\begin{figure}[!tt]
	\centering
	\includegraphics[width=\columnwidth]{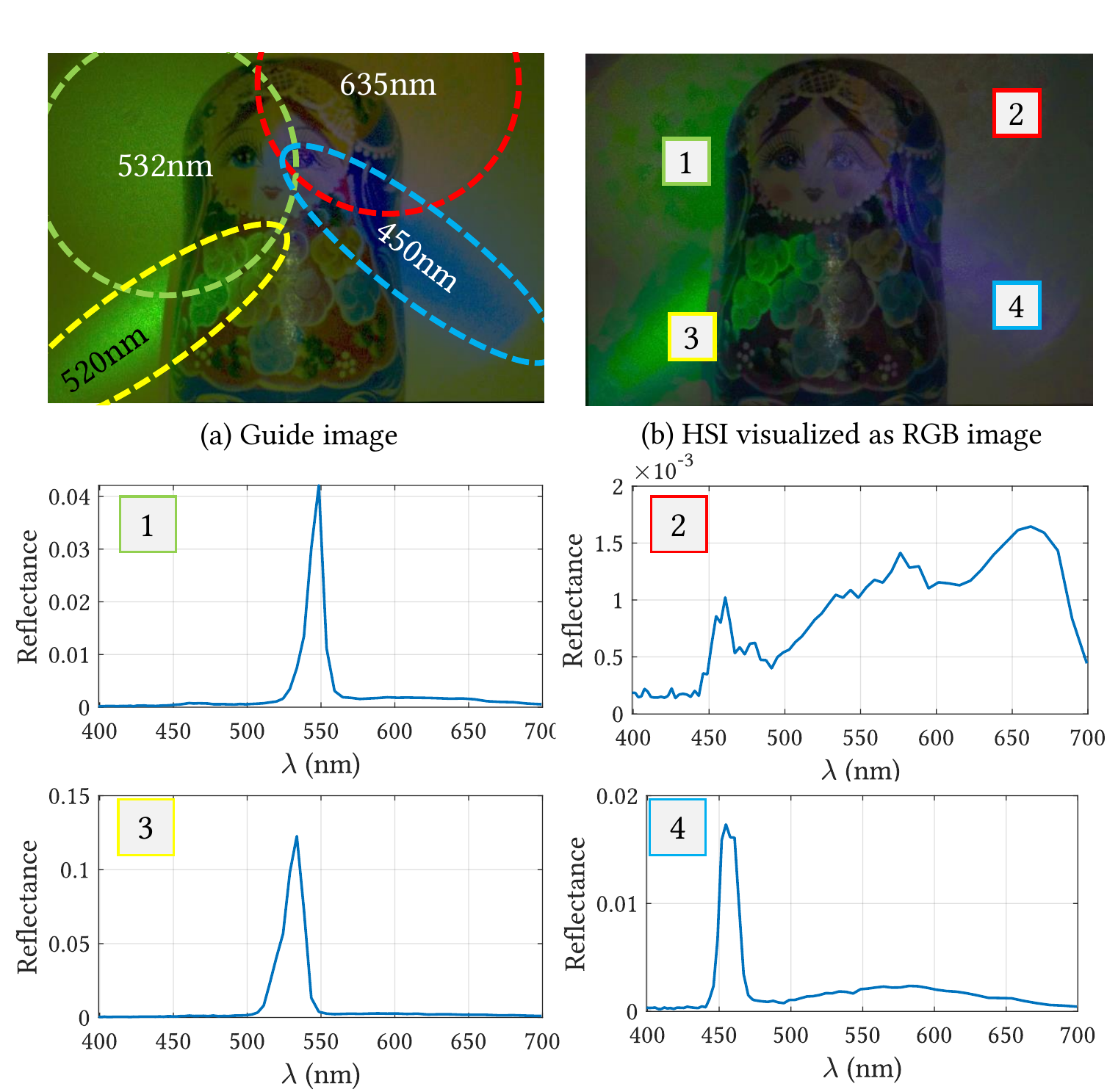}
	\caption{\textbf{Scenes with complex illumination.} Our sampling and reconstruction approach does not require any assumptions on smoothness of scene spectra. To test this, we (a) illuminated a scene with four lasers and a white LED light and then reconstructed the HSI with a our approach. The RGB image is shown in (b) and the spectral profiles marked at four locations can be seen below the images. The laser profiles are distinctly visible. the mixture of illuminants is visible in points 2 and 4. }
	\label{fig:complex_illum}
\end{figure}

\bpara{Complex illumination.}
Since our setup does not rely on smoothness of spectral profiles, but directly samples them in the scene, we can capture HSIs with arbitrarily complex spectrum.
To illustrate this, we illuminated a scene consisting of a doll with four lasers of wavelengths $450, 520, 532$ and $635$ nm and had a global illumination of LED light source. We then captured an HSI with our method. 
Figure \ref{fig:complex_illum} shows the scene with location of lasers, as well as spectral profiles at four select points.
Our setup localizes arbitrarily sharp spectra and also is capable of estimating mixed illumination, as can be seen in spectrum at points 2 and 4, where there is a mix of laser and LED light sources.

\begin{figure}[!tt]
	\centering
	\includegraphics[width=\columnwidth]{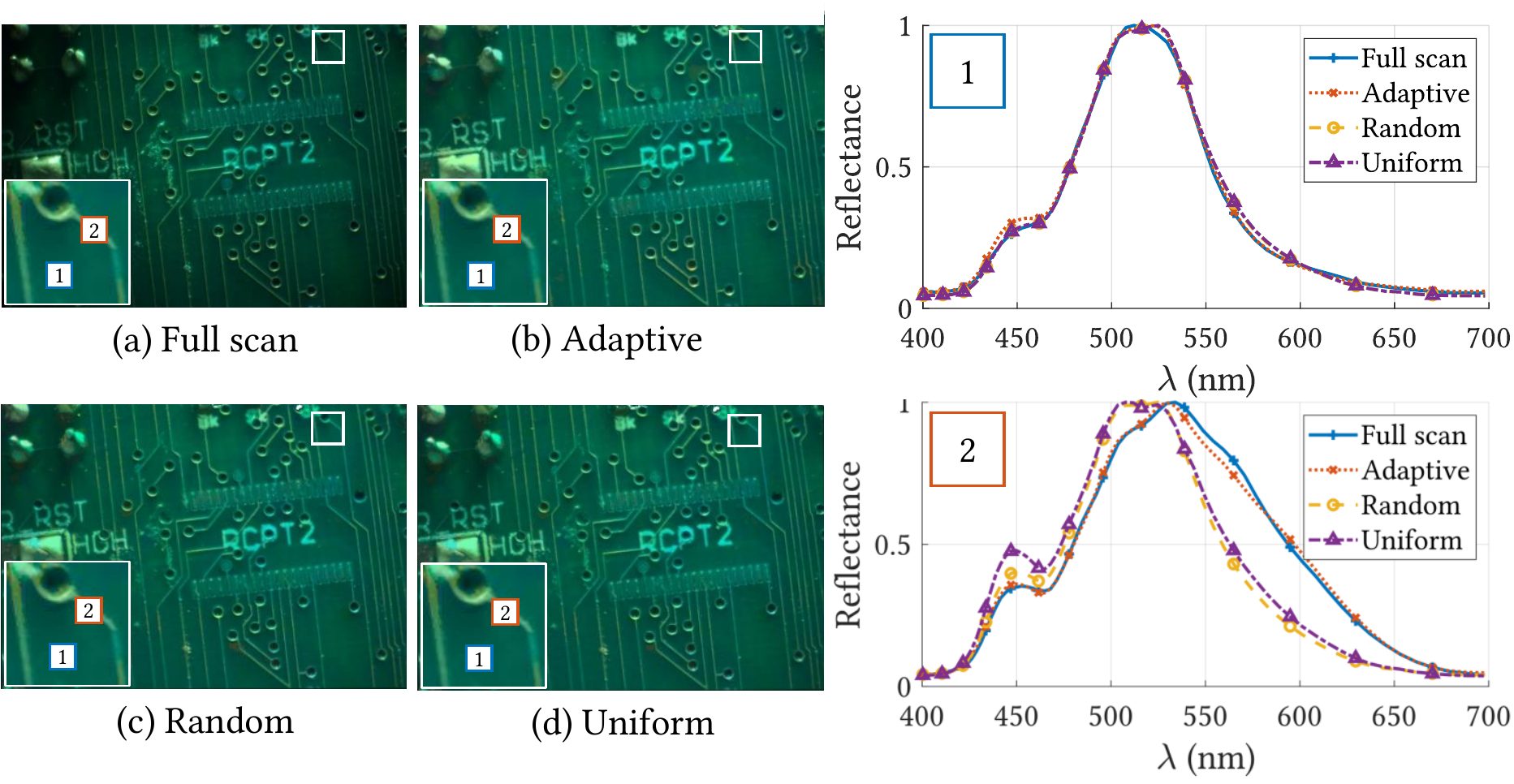}
	\caption{\textbf{Comparison of sampling strategies.} The benefits of adaptation are more evident when the scene has complex texture. We imaged a printed circuit board that consists of very thin metallic tracks. Due to adaptive sampling, our method ensures that at least one pixel is sampled on the metallic track. In contrast, applying either random or uniform sampling misses the track. This results in spectrum being the same for point 1 on the background and point 2 on the track.}
	\label{fig:real_sampling_method}
\end{figure}

\bpara{Effect of sampling strategy.}
Adapting the pattern on SLM allows us to sense intricate structures in the scene. To test this, we imaged the HSI of a printed circuit board (PCB) that consisted of thin metal tracks.
We sampled the scene using three different patterns, namely an adaptive pattern (proposed), a random pattern such that there was no spectral overlap, and a uniformly spaced pattern which also ensured that there was no spectral overlap.
We then constructed using our rank-1 approximation strategy. The images and spectrum at a selected location is shown in Fig. \ref{fig:real_sampling_method}.
The spectral profiles at the two marked points show that adaptivity allows us to separate the metal track from background, whereas a fixed pattern (random or uniform) will miss sampling the metal track.

\begin{figure}[!tt]
	\centering
	\includegraphics[width=\columnwidth]{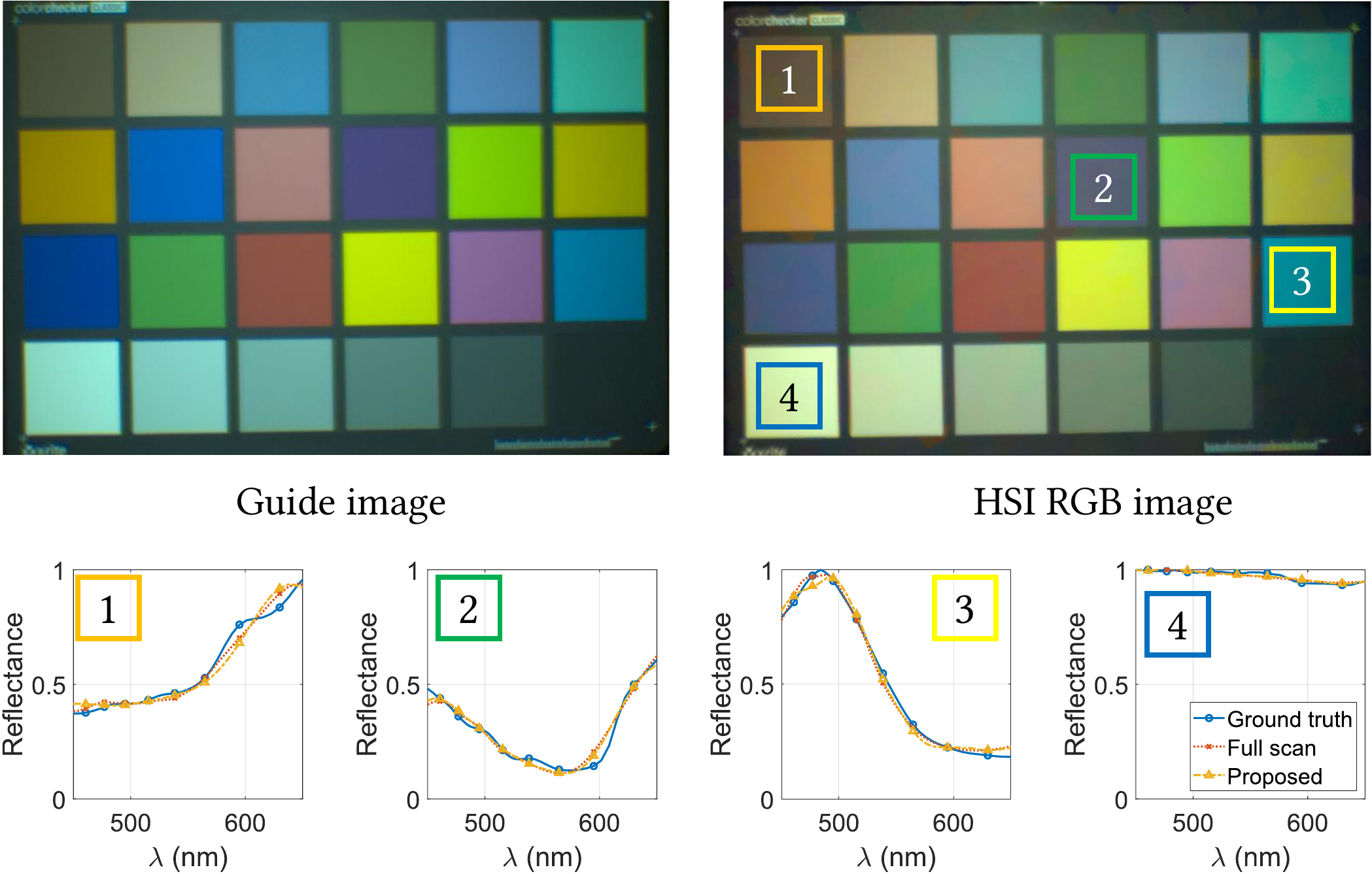}
	\caption{\textbf{Capturing Macbeth data.} We captured a Macbeth color chart and compared against reference as well as spectral profiles captured by fully scanning the HSI. }
	\label{fig:macbeth}
\end{figure}

\bpara{Challenging scenarios.}
We mentioned in the paper that our approach may not give accurate results when superpixellation does not separate regions of different spectra.
An example is shown in the last row of Fig. \ref{fig:sup_real_master}, of a microscopic specimen of stem of Nymphae.
The spectra shown in the last column as point 1 does not match the ground truth, shown using a red mark in the zoomed in picture.
One way would be to capture multiple measurements, with each measurement focused on measuring areas that are less sampled.
This would make an interesting future direction for our approach.
\end{appendices}

{
	\small
\bibliographystyle{IEEEtran}
	\bibliography{refs}
}

\ifpeerreview \else






\fi

\end{document}